\documentclass[a4paper,fleqn]{cas-sc}
\usepackage{stfloats}
\usepackage{subfiles}
\usepackage[caption=false,font=footnotesize]{subfig}
\usepackage[export]{adjustbox} 
\usepackage{xcolor}

\usepackage[normalem]{ulem}\let\printorcid\relax

\usepackage[utf8]{inputenc} 
\usepackage[english]{babel}
\usepackage{amsmath,amsfonts,amssymb,accents}
\usepackage{mathtools}                  
\usepackage{bbm,bm} 
\usepackage{graphicx} 
\usepackage{hyperref}

\usepackage{outlines}
\usepackage{pgfplots}
\usepackage{tikz}
\usetikzlibrary{matrix,shapes,decorations.pathreplacing,fit,backgrounds}
\usetikzlibrary{plotmarks}
\usetikzlibrary{arrows.meta}
\usepackage{tikz-3dplot}
\pgfplotsset{compat = newest}
\usepgfplotslibrary{patchplots}
\usetikzlibrary{calc}
\pgfplotsset{
      axis line style={black},
      every axis label/.append style ={black},
      every tick label/.append style={black}  
    }

\usepackage{hyperref}
\usepackage[capitalize,nameinlink]{cleveref}       
\newcommand{\inD}{\quad \qquad \text{in}\;{\phantom{\partial}}\Omega}

\newcommand{\dOmega}{\mkern 2mu\mathrm{d}\Omega}
\newcommand{\numberset}[1]{\bm{#1}}
\renewcommand{\r}{\numberset{r}}
\usepackage{siunitx}
\usepackage[numbers]{natbib}

\makeatletter
\def\input@path{{src/}
                {}}
\makeatother
\graphicspath{{img/}}

\newcommand{\software}[1]{%
  \begingroup
  \fontfamily{pcr}\selectfont
  \fontsize{9pt}{10pt}\selectfont
  \StrLeft{#1}{1}[\firstchar]%
  \StrGobbleLeft{#1}{1}[\rest]%
  {\fontsize{10pt}{11pt}\selectfont \uppercase{\firstchar}}%
  \uppercase{\rest}%
  \endgroup
} 
\usepackage{listings}
\usepackage{inconsolata}
\newcommand{\sculpt}[1]{{\fontfamily{zi4}\selectfont\fontsize{9}{11}\selectfont\spaceskip=0.2em plus 0.2em minus 0.2em #1}}

\definecolor{lightgray}{HTML}{EEEEEE}
\definecolor{lightyellow}{HTML}{FFF2CC}
\definecolor{lightblue}{HTML}{C9DAF8}
\definecolor{lightgreen}{HTML}{D9EAD3}

\let\printorcid\relax

\begin{document}
\shorttitle{Feasibility of spectral-element modeling of wave propagation through the anatomy of marine mammals}
\shortauthors{C. García A. \textit{et~al.}}
\title [mode = title]{Feasibility of spectral-element modeling of wave propagation through the anatomy of marine mammals}
\author[1]{Carlos Garc\'\i a~A.}[]
\author[2]{Vladimiro Boselli}
\author[3]{Aida Hejazi Nooghabi}
\author[4]{Andrea Colombi}
\author[1]{Lapo Boschi}
\address[1]{Dipartimento di Geoscienze, Universit\`a degli Studi di Padova, 35131 Padova, Italy}
\address[2]{Institute for Electromagnetic Sensing of the Environment, National Research Council of Italy CNR-IREA, Milano, 35122, Italy}
\address[3]{Institute of Geophysics, Centre for Earth System Research and Sustainability (CEN), Universität Hamburg, Hamburg, 20146, Germany}
\address[4]{School of Architecture, Design and Civil Engineering, ZHAW Zürich University of Applied Sciences, Zürich, 8401, Switzerland}

\begin{abstract}
This study introduces the first 3D spectral-element method (SEM) simulation of ultrasonic wave propagation in a bottlenose dolphin (\textit{Tursiops truncatus}) head. Unlike traditional finite-element methods (FEM), which struggle with high-frequency simulations due to costly linear-system inversions and slower convergence, SEM offers exponential convergence and efficient parallel computation. Using Computed Tomography (CT) scan data, we developed a detailed hexahedral mesh capturing complex anatomical features, such as acoustic fats and jaws. Our simulations of plane and spherical waves confirm SEM’s effectiveness for ultrasonic time-domain modeling. This approach opens new avenues for marine biology, contributing to research in echolocation, the impacts of anthropogenic marine noise pollution and the biophysics of hearing and click generation in marine mammals. By overcoming FEM's limitations, SEM provides a powerful scalable tool to test hypotheses about dolphin bioacoustics, with significant implications for conservation and understanding marine mammal auditory systems under increasing environmental challenges.
\end{abstract}    

\maketitle


\section{\label{sec:intro} Introduction}

Numerical simulations of wave propagation in biological media have garnered significant attention in recent years, with a vast span of applications, ranging from medical physics \cite{grimal_finite_2004,vantwout_fast_2015,gu_modeling_2015,bachmann_source_2020} to bioacoustics \cite{hejazinooghabi_contribution_2021, wei_validated_2024} and ecology \cite{castellazzi_assessment_2012}. Some of the authors of this study have pointed out, in earlier work, the difficulty of explaining the accuracy of auditory source localization by dolphins, particularly if one considers that they have no pinnae: understanding how elastic waves propagate through their bodies will help us unraveling some of the mysteries of biosonar \cite{hejazinooghabi_contribution_2021,nachtigall_biosonar_2016}. 

Until recently, the feasibility of time-domain simulations through realistically complex biological structures remained uncertain. Advancements in computational power and a growing interest in biodiversity conservation have driven greater focus on numerical methods. These methods are developed, adapted, or applied to study, for instance, sound generation and propagation in animals, especially when experiments are not feasible \cite{kazemi_realistic_2023, feng_biosonar_2019}. The leading numerical solvers in this context include the finite-element method (FEM) \cite{comsolmultiphysics_comsol_2024,alnaes_fenics_2015,hecht_new_2012,arndt_dealii_2021}, the boundary element method (BEM) \cite{haqshenas_fast_2021,betcke_bemppcl_2021}, ray tracing \cite{porter_gaussian_1987}, and pseudospectral methods \cite{treeby_modelling_2014}. While each approach has specific strengths and weaknesses, performing wave-based simulations that account for fluid-elastic coupling, operate in the ultrasonic frequency range, and solve time-domain propagation has historically been highly impractical. This is where the spectral-element Method (SEM) \cite{patera_spectral_1984,komatitsch_introduction_1999} stands out. SEM inherits from FEM the capability to handle complex geometries, but, unlike FEM, requires no linear-system inversion, which is a notoriously costly endeavor \cite{komatitsch_wave_2000}. In addition, compared to traditional FEM, SEM converges exponentially rather than with second-order accuracy and as such it is particularly efficient for solving time-domain wave propagation problems. Well known and widely used examples of SEM solvers are the commercial package \software{SALVUS} \cite{afanasiev_flexible_2018}  and its open-source counterpart \software{SPECFEM} \cite{komatitsch_spectralelement_2002,komatitsch_spectralelement_2002a}, employed in this study. \software{SPECFEM} is \software{FORTRAN}-based, with no user interface, but relatively simple to use and very efficiently parallelized. It was originally designed for use on large supercomputer clusters, where it still excels. 

\software{SPECFEM} was originally developed by seismologists for global-scale seismology applications. In over two decades, it has been used successfully by geophysicists all over the world; its effectiveness has been largely proved at scales of (at least) a few km and in the infrasound frequency range. Scaling it to higher frequencies and shorter spatial scale lengths is, in theory, straightforward. However, the anatomy of marine mammals presents complexities not found in geological media. Reducing such intricate structures to discrete, numerical “meshes” turns out to be a non-trivial problem, to say the least. As a result and despite their potential, \software{SPECFEM} and SEM in general, have not yet been applied to simulations of wave propagation in marine mammals. The literature on biological application of SEM is, in fact, sparse (e.g., \cite{marty_fullwaveform_2022,marty_elastic_2022,marty_transcranial_2024}). Perhaps the most significant challenge is that SEM inherently relies on hexahedral element meshes (see \cite{ferroni_discontinuous_2017} to understand the difficulties of applying SEM to non-hexahedral meshes). Unlike tetrahedral meshes, which can be generated algorithmically e.g. via Delaunay triangulation, there is no automated tool to generate hexahedral meshes: such an algorithm is often referred to as the ``Holy Grail'' of mesh generation \cite{blacker_meeting_2000}. Potential users of SEM are cautioned that, as the apocryphal rule of thumb states, they may spend around 80\% of their time creating valid models and meshes before those can be fed to the SEM solver itself. On the other hand, other numerical methods such as traditional FEM, while easier to set up, often face scalability challenges, as they require that linear systems be solved at each numerical iteration. Seismology applications of SEM suggest that mesh-related issues can be overcome, and we infer that SEM is potentially the most powerful available tool for addressing 3D wave propagation in biological structures. This is particularly true for marine mammals, which have complex and yet to be fully explored auditory systems. Indeed, the need to model intrinsically complex shapes, to identify the contribution of different anatomical features, to place virtual sensors in otherwise inaccessible anatomical locations and to estimate the effect of potentially harmful vibrations, all emphasize the importance of efficient numerical simulations.

Several studies, mostly carried out by Chong Wei and his team, have addressed the problem of simulating sound propagation in marine mammals, especially in dolphins. Various hypotheses have been proposed to explain how dolphins perceive sound \cite{wei_finite_2018,wei_modeling_2020}, involving the roles of the melon \cite{zhang_directional_2017}, forehead shape \cite{wei_distinctive_2022} and other anatomical structures \cite{wei_role_2016,song_inducing_2016}; the biosonar beam formation \cite{wei_numerical_2018} and even biosonar-inspired metamaterials \cite{dong_physical_2019}. These studies have relied on 2D models, which, while they do contribute to advance our understanding of dolphin hearing and biosonar, do not capture its full complexity. The recent 3D FEM-based model of \citet{wei_validated_2024} is very promising for overcoming this difficulty, but we submit that, in the long run, SEM might provide us with an equally or more efficient tool for 3D numerical modeling.

In this paper, we evaluate the feasibility of SEM-based simulations of 3D time-domain wave propagation in a real bottlenose dolphin head (\textit{Tursiops truncatus}). We outline the challenges of using SEM for such simulations in animal tissues and anatomical structures, detailing the modeling process from input data derived from CT scans, to geometry creation, mesh generation, and finally, wave propagation based on the \software{SPECFEM3D} module of \software{SPECFEM}.

\section{\label{sec:theory} The Spectral-Element Method: General notions}

All simulations in this study were conducted using the \software{SPECFEM3D} package \cite{komatitsch_spectral_1998,komatitsch_introduction_1999}, a widely known and validated implementation of SEM. SEM \cite{patera_spectral_1984} is a so-called high-order finite-element method, i.e., it uses high-order polynomial basis functions to approximate the numerical solution within each element of a discrete mesh. A brief overview of the method is presented here, intended to highlight the aspects of SEM that are most relevant to our application. For an accessible introduction to SEM, compared to other numerical methods, the reader is referred to \citet[Ch.\ 7]{igel_computational_2016}. 

Consider the scalar potential $\varphi(\bf{x}$$, t)$ and a differential operator $\mathcal{L}$ describing the propagation of acoustic or elastic waves excited by a force $f$ in lossless bounded acoustic and elastic domains $\Omega_{a}$ and $\Omega_{e}$, respectively, with $\Omega_a\cup\Omega_e=\Omega$.

\begin{equation}\label{eq:Lu_f}
  \mathcal{L}\varphi = f \inD_{a;e}.
\end{equation}
To solve \cref{eq:Lu_f}, we first derive the weak form by multiplying the operator $\mathcal{L}$ by a test function $v \in V$ (a suitable function space) and integrate over the domain. The problem is then reformulated as finding $\varphi \in V$ such that

\begin{equation}\label{eq:Lu_f_weak}
\int_{\Omega_{a;e}} v \mathcal{L}\varphi \, \dOmega_{a;e} = \int_{\Omega_{a;e}} v f \, \dOmega_{a;e}, \quad \forall v \in V.
\end{equation}
The fluid-elastic coupled system in \cref{eq:Lu_f_weak} can be expressed in matrix form \cite{marty_transcranial_2024}

\begin{equation}\label{eq:matrix_form}
\left[\begin{array}{c c}
        {\bf M}_a \partial t^2 + {\bf K}_a & -{\bf C}_a\\
        -{\bf C}_e\partial t^2 & {\bf M}_e \partial t^2 + {\bf K}_e
 \end{array}\right]
\left(  \begin{array}{c}
        {\bf \varphi}\\ 
        {\bf u}
\end{array}\right)
        =
\left(  \begin{array}{c}
        {\bf F}_a\\ {\bf F}_e
        \end{array}\right),
\end{equation}
where $\bf{u}(\bf{x}$$, t)=\rho^{-1}\nabla \varphi$ denotes the displacement of a moving particle, $\rho$ the volumetric density, ${\bf M}_{a;e}$ the mass matrix, ${\bf K}_{a;e}$ the stiffness matrix, ${\bf C}_{a;e}$ the coupling matrix and ${\bf F}_{a;e}$ is the source term expressed in the right-hand side of \cref{eq:Lu_f_weak}. The attractiveness of the SEM is that the global mass matrix is diagonal by construction, therefore solving the coupled fluid-elastic system, i.e. \cref{eq:matrix_form} can be implemented with matrix-inversion-free time-domain solvers and explicit time-stepping schemes (e.g., Newmark).

\subsection{Numerical Parametrization of the Solution}
Within each discrete non-overlapping element, the solution $\varphi$ is approximated by a high-order polynomial of degree $N$

\begin{equation}
\varphi(\mathbf{x}) = \sum_{i=0}^{N}  \alpha_i \phi_i(\mathbf{x}).
\end{equation}
The basis functions $\phi_i(\mathbf{x})$ are defined at the Gauss-Legendre-Lobatto (GLL) points within the element, and scaled by the nodal coefficients $\alpha_i$. Namely,

\begin{equation}
\phi_i(\xi) = \prod_{\substack{j=0 \\ j \neq i}}^{N} \frac{\xi - \xi_j}{\xi_i - \xi_j}.
\end{equation}

\subsection{Fluid-elastic coupling}

When the medium to be modeled includes both fluid and solid regions, \software{SPECFEM} implements the full elastic equation only in the solid elements and the more compact acoustic wave equation in the fluid ones. This reduces the computational cost of the simulation, as the acoustic wave equation is less computationally expensive (scalar) than the elastic wave equation (vector) \cite{komatitsch_wave_2000}. At fluid-solid interfaces, boundary conditions are implemented explicitly, prescribing the continuity of pressure and normal stress.

\section{\label{sec:methods}From Computed Tomography Scans to Numerical Simulations}
\begin{figure}
\centering
\subfloat{
  \includegraphics[width=8cm]{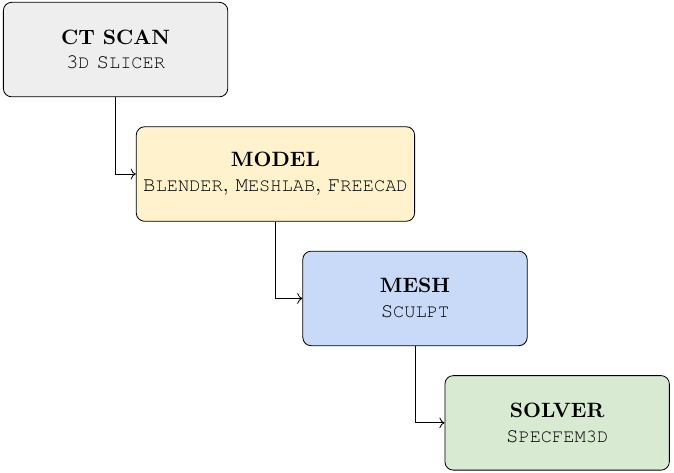}}
    \caption{\label{fig:workflow}Workflow. Software packages used at each step are specified in small fonts in the corresponding block.}
\end{figure}

\software{SPECFEM} requires that the medium to be modeled be projected onto a hexahedral mesh and creating hexahedral meshes is notoriously complex and challenging. Successful simulations depend on each preceding step. To make sure that our results are fully reproducible and to facilitate future applications of our method, we next describe the entire procedure (summarized by \cref{fig:workflow}) in some detail.

\subsection{\label{subsec:ct_scans}Computed Tomography Scans}
\begin{figure}
    \centering
    \subfloat[\label{fig:slicer_3d}]{
        \includegraphics[width=0.4\textwidth]{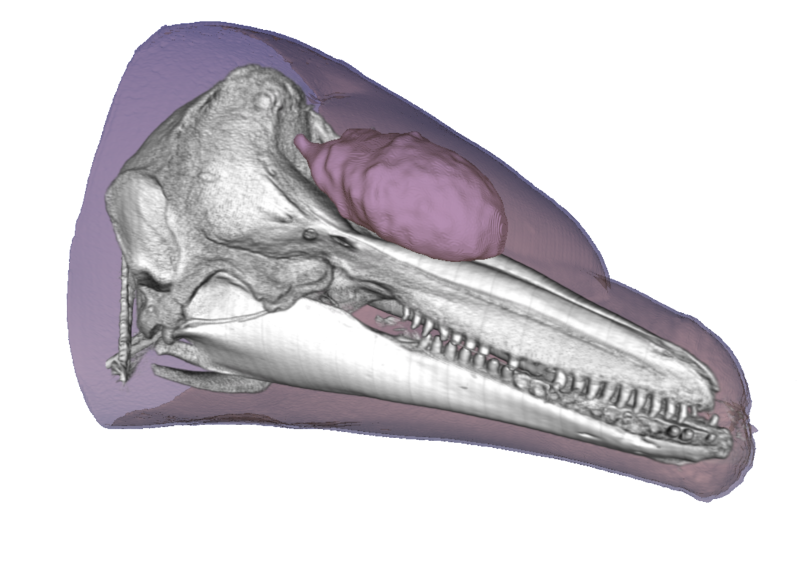}}\\
    \subfloat[\label{fig:slicer_cuts}]{%
      \includegraphics[width=0.8\textwidth]{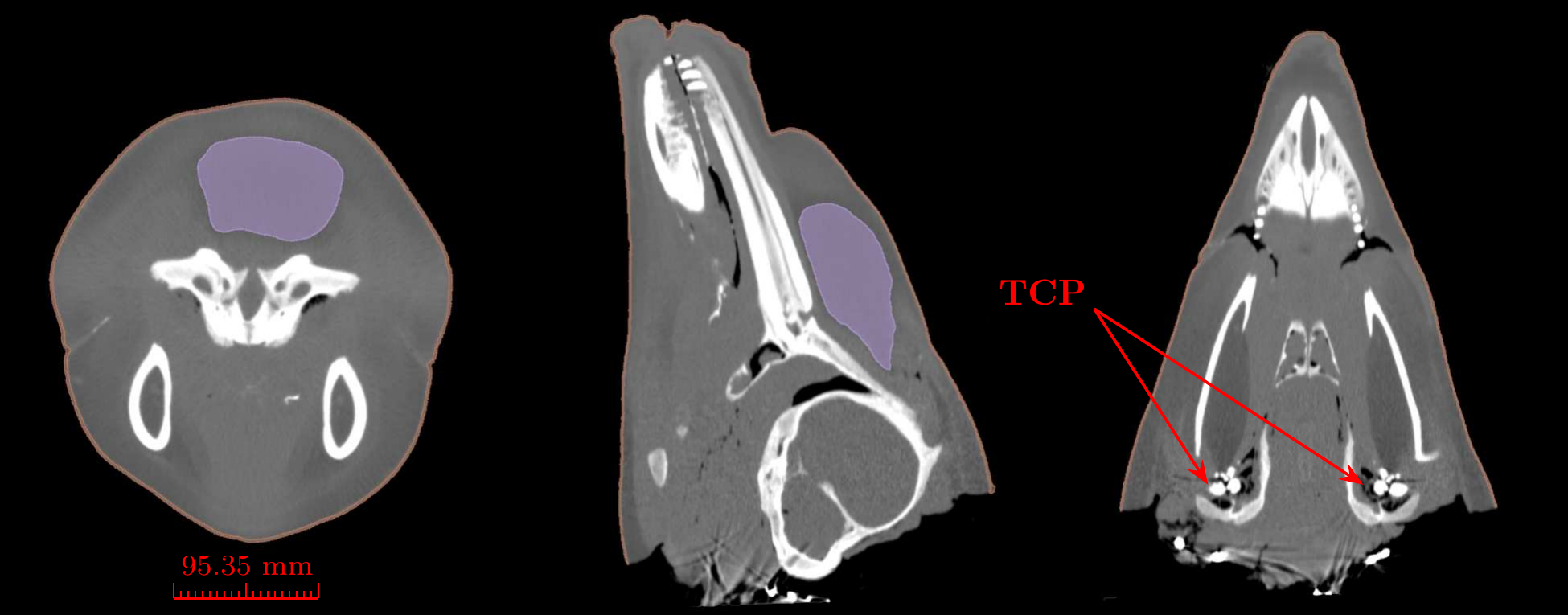}}
    \caption{\label{fig:slicer}Segmentation in \protect\software{3D} \protect\software{SLICER}. (a) Original skull before segmentation. Melon and skin boundary after segmentation. Bones were segmented independently and subsequently merged. (b) The frontal (left), side (centre) and bottom (right) views illustrate the segmentation processes for the skin (thin orange lines) and the melon (purple area). The Tympano-Periotic Complex (TPC) is visible, although the CT scan quality is not sufficient for high-detail segmentation.}
\end{figure}

To create a 3D model of the dolphin head, we used Computed Tomography (CT) scans of a bottlenose dolphin (\textit{Tursiops truncatus}) obtained from the Department of Comparative Biomedicine and Food Science at the University of Padua. The resolution of the CT scan was $0.813 \times 0.813 \times 1.5$ mm. The head corresponds to an old individual with some missing teeth, as shown in \cref{fig:slicer_3d}. The CT scans were processed using \software{3D} \software{SLICER} \cite{fedorov_3d_2012}, an open-source software for medical image processing and three-dimensional visualization. CT scans are commonly formatted as multiple DICOM files; each file is a 2D image slice in which each pixel represents a voxel in 3D space. The intensity of each voxel corresponds to the Hounsfield unit (HU) value of each tissue \cite{brooks_quantitative_1977}. We relied on HU values to differentiate between different types of tissues, such as bone, fat, and soft tissue. We employed \software{3D} \software{SLICER} to identify and isolate the different types of tissues found in the dolphin, through a process usually referred to as ``segmentation''. This involved manually selecting the regions of interest and applying thresholding techniques (see \cref{fig:slicer}). 
We segmented four types of tissue: bone, acoustic fat, melon, and soft tissue. The soft tissue is bounded externally by the skin (see \cref{fig:slicer_cuts}) and internally by the remaining tissues. Bones were segmented independently and then combined to form the skull and lower jaws. Once the segmentation was complete, we exported each segmented part in the stereo-lithography format (STL). STL is a widely used format for 3D printing and computer-aided design (CAD) applications. It represents the outer surface of a volume using a mesh of triangular facets.

\begin{figure}
    \centering
    \subfloat[-\label{fig:model_a}]{
        \includegraphics[width=6cm]{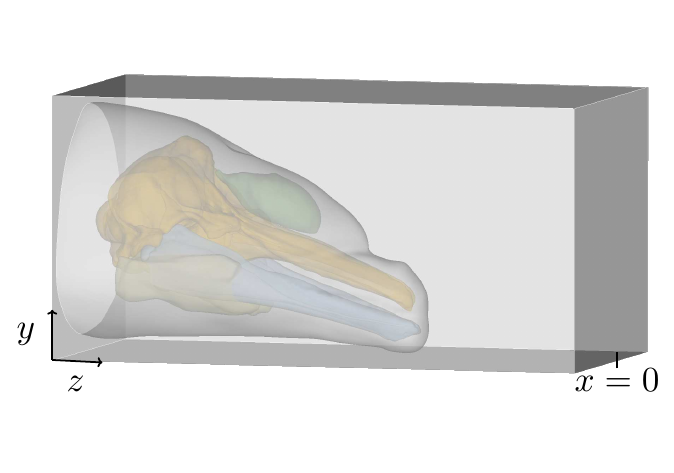}}\hspace{2cm}
    \subfloat[\label{fig:model_b}]{ 
        \includegraphics[width=6cm]{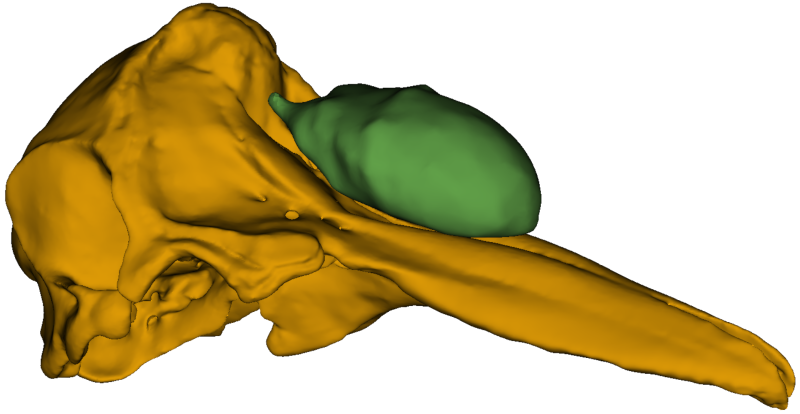}}\\
    \subfloat[\label{fig:model_c}]{
        \includegraphics[width=5cm]{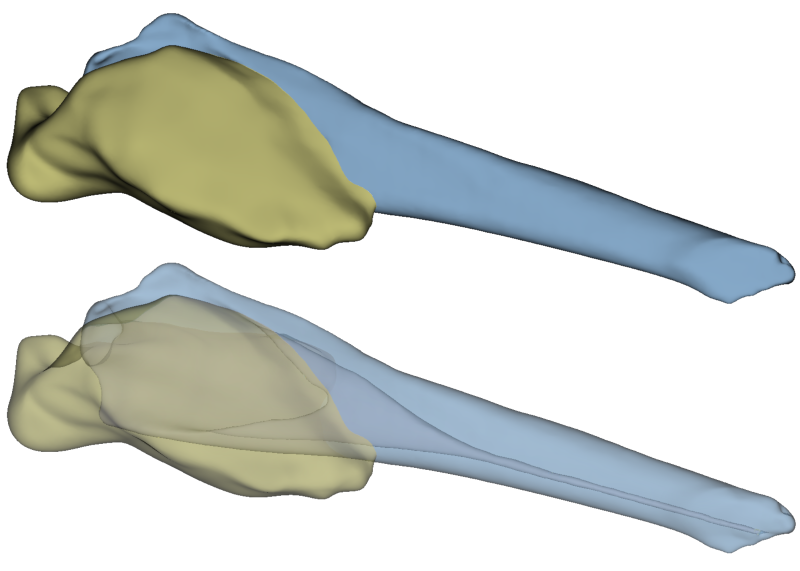}}\hspace{4cm}
    \subfloat[\label{fig:model_d}]{
        \includegraphics[width=2.5cm]{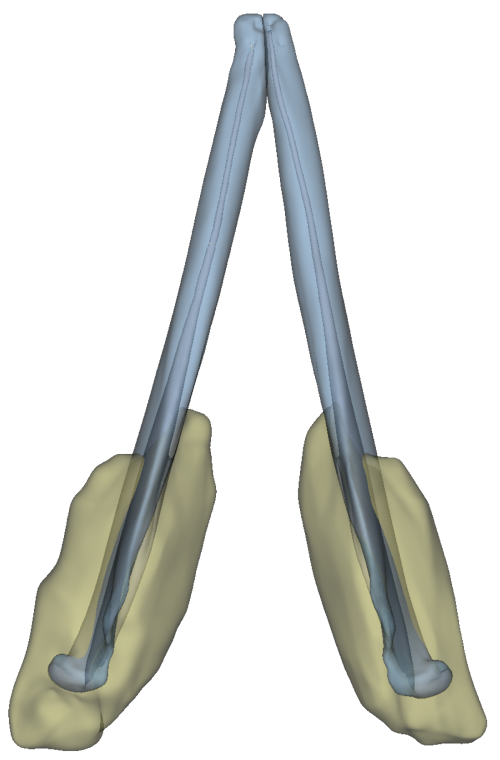}}
    \caption{\label{fig:model}Model geometry. (a) Complete model: the shaded box corresponds to the volume of water surrounding the dolphin head. Other anatomical features depicted are: the melon (green), skull (orange), lower jaws (light blue) and acoustic fats (light yellow). (b) Skull and melon geometry. (c) Side view: melon, jaws and acoustic fats, slightly transparent (top) and opaque (bottom). (d)  Top view: jaws and acoustic fats (slightly transparent). Transparency is necessary to visualize the bones that the fats partly enclose.}
\end{figure}

\subsection{\label{subsec:geometry}Geometry manipulation}

The resulting STL files were imported into \software{BLENDER} \cite{blenderfoundation_blender_2022}, where we performed some geometry modifications, such as smoothing sharp edges, removing all teeth and filling some internal canals within the skull, as well as some foramen holes. This helped simplify the subsequent meshing process.
In fact, it would have been possible, in principle, to keep much of the details that we decided to smooth away. However, this would have implied a sharp decrease in the mesh size, with corresponding increase in computational cost, which seemed unnecessary to our current feasibility study (conducted on relatively limited computational resources).

We next refined the mesh in \software{MESHLAB} \cite{cignoni_meshlab_2008}, removing any duplicate vertices, fixing non-manifold edges and other artefacts, occasionally left by \software{BLENDER}, that could cause problems in the subsequent process. This step can be implemented automatically via the \software{PYTHON} API \software{PYMESHLAB} \cite{muntoni_pymeshlab_2021}, for all the files of interest at once. There are also multiple purchasable plug-ins in \software{BLENDER} intended to do this, but we have not explored them yet.

Next, the resulting individual STL files were imported into \software{FREECAD} \cite{rieger_freecad_2001}. A complete 3D model was constructed, including a water-filled rectangular box surrounding the dolphin head (\cref{fig:model_a}), with dimensions $0.42 \times 0.42 \times 0.86$ m. The model was designed to be large enough to capture the incoming signal well and small enough to minimize the computational cost of the simulation. The head had been chopped before CT scans were taken so some smoothing was needed. In \software{SPECFEM3D}, outer surfaces must be aligned with the Cartesian plane for imposing non-reflection boundary conditions. The back of the head was sculpted and sliced to align with the origin $xy$-plane, as shown in \cref{fig:model_a}.  

The imported STL files represent the outer surface mesh of a 3D volume. The next step consists of explicitly create all interfaces between different domains, e.g., between soft and fat tissues, to ensure continuity of nodes when finally building the hexahedral mesh. The easiest way to achieve this is via so-called ``boolean operations'', i.e., subtracting, intersecting, and merging different objects that are in contact with each other. \software{FREECAD} provides simple but powerful tools for such operations. Specifically, we used the ``boolean fragments'' operation between objects that share surfaces, starting from the inner tissues and progressing to the outer ones.  Boolean fragments create additional nodes when performing boolean operations, ensuring that meshes are conformal \cite{rieger_freecad_2001}. \software{BLENDER} and \software{MESHLAB} can also perform boolean operations, but they do not incorporate ``boolean fragments'' and the resulting meshes are not conformal. 

Acoustic fats and jaws (see \cref{fig:model_c,fig:model_d})  are enclosed within and in contact with the soft tissues. To properly honor such a complex structure, we first create an auxiliary object from the union of the acoustic fats and the jaws. This union was performed as a boolean fragment. We then subtract this auxiliary object from the fat tissues, which ensures explicit nodes at both boundaries. Creating the boundary surfaces for the skull and the melon (e.g. \cref{fig:model_b}) is simpler and can be done directly, without any further auxiliary object.

Last, the back of the head is aligned with the Cartesian plane $xy$. At the same position in $z$, we place the back side of the water volume. Care must be taken when performing the boolean subtraction between the soft tissues and the water object, as most algorithms do not handle coplanar faces well. To avoid issues, we slightly moved one object, the head in this case, a very small distance, e.g. 1 mm in the $z$ direction. We then performed the subtraction and, finally, took away from the head the remaining 1 mm section.

\subsection{\label{subsec:meshing}Meshing process}

\begin{figure}
    \centering
    \subfloat[\label{fig:mesh_n}]{
        \includegraphics[height=4.5cm]{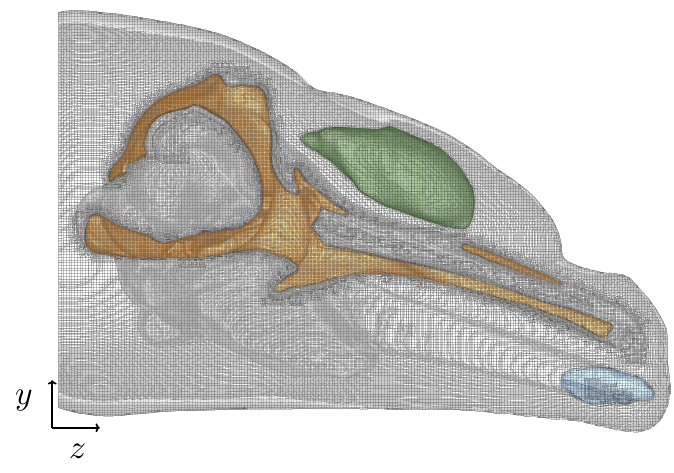}}\hspace{1cm}
    \subfloat[\label{fig:mesh_a}]{
        \includegraphics[height=4.5cm]{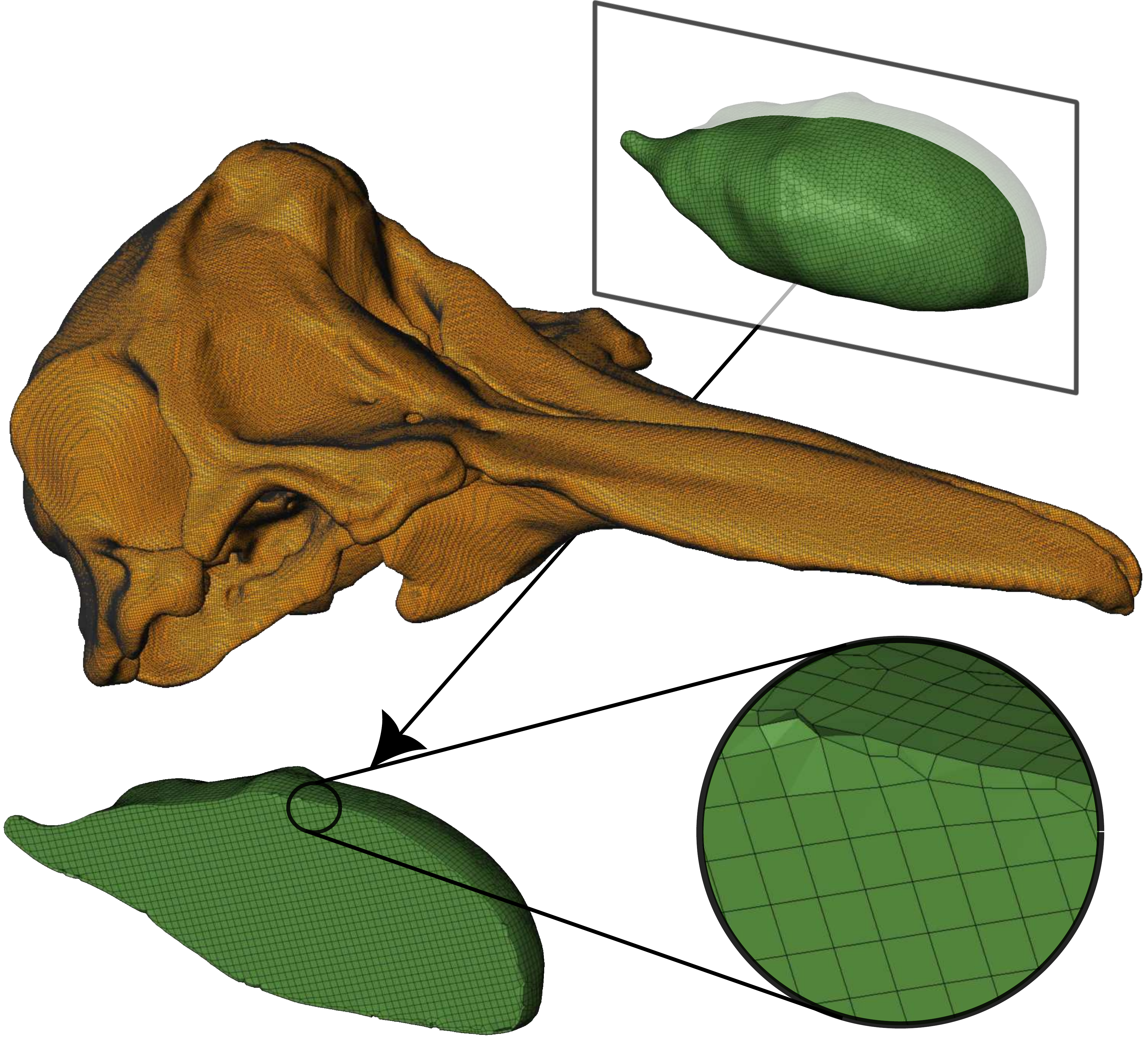}}\\
    \subfloat[\label{fig:mesh_b}]{
        \includegraphics[height=4.5cm]{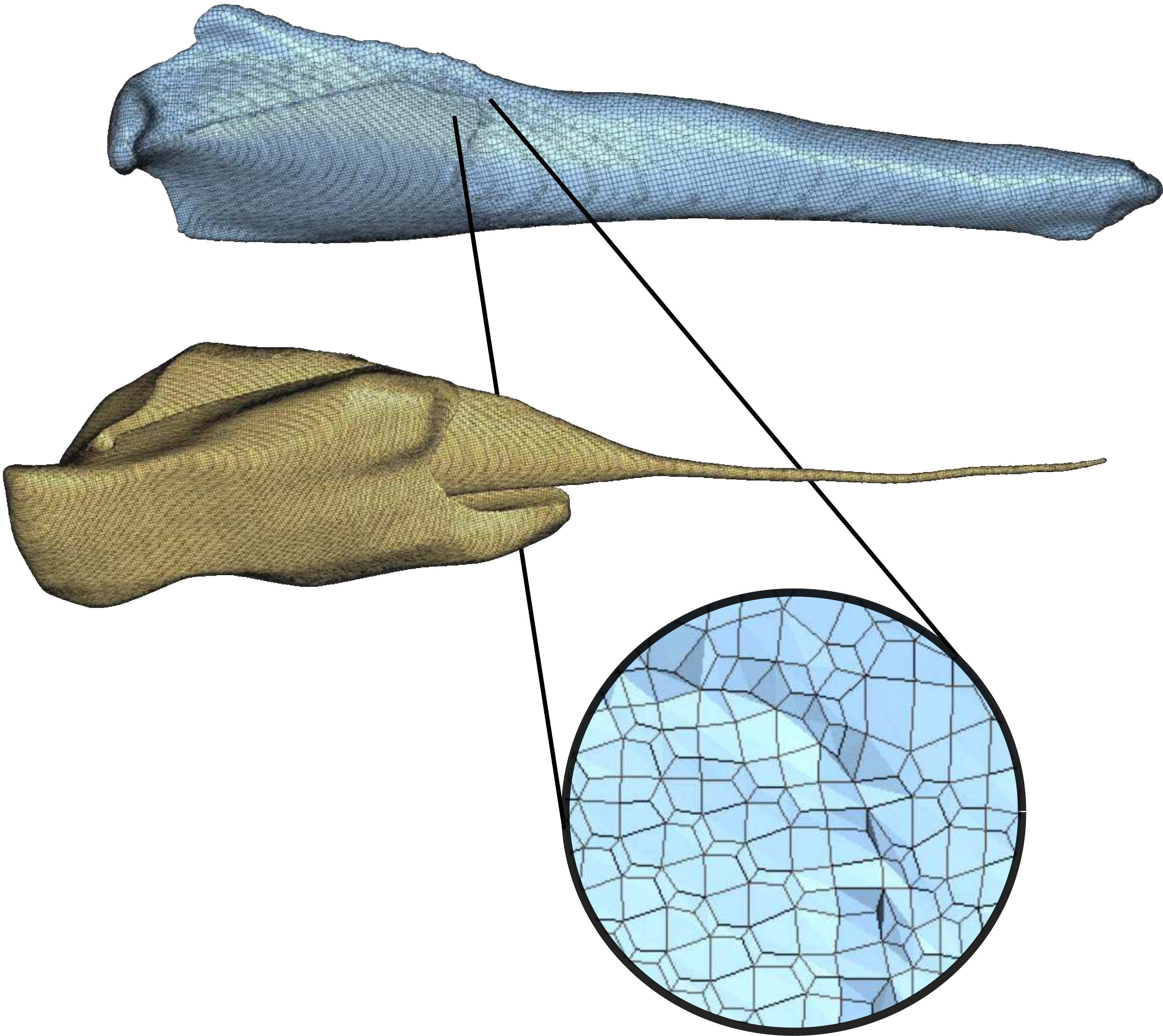}}\hspace{2.5cm}
    \subfloat[\label{fig:mesh_c}]{
        \includegraphics[height=4.5cm]{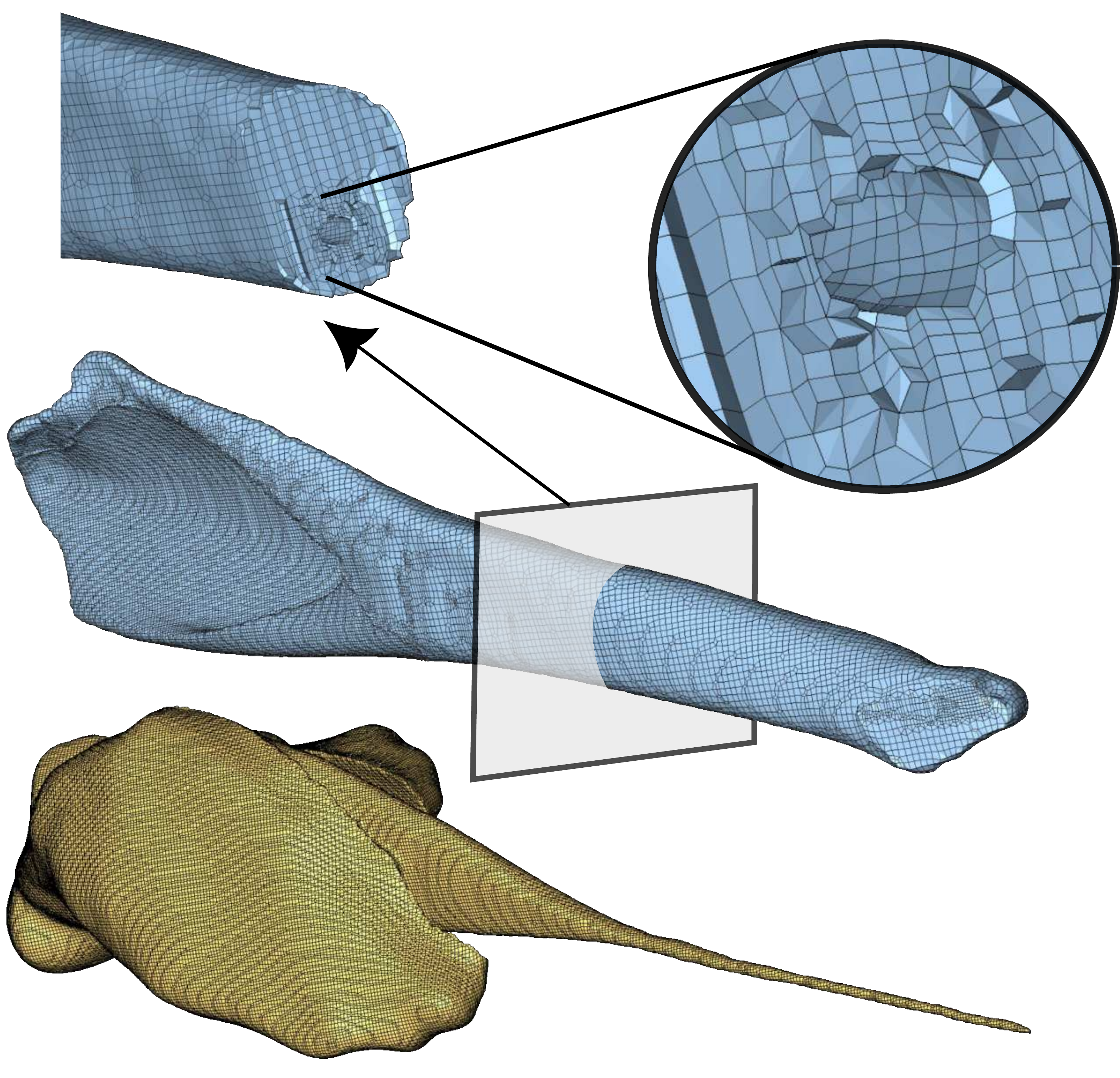}}
    \caption{\label{fig:mesh}a) Transparent view of the mesh. A clipping plane along the median $yz$-plane at $x=0$ was applied to reveal its  internal structure. b) Details of the skull and melon. Most of the interior elements are cuboids with their external faces conforming to the interface geometry, as shown in the enlarged view outlined by a black circle.  c) Right side of the jaw and acoustic fats. The detail shows the conformal mesh between both tissues. d) Left side of the jaw and acoustic fats. The interior bone cavity, filled with fat tissue and shown in detail, is a good example of mesh adaptivity.}
\end{figure}

\begin{table}
  \begin{tabular}{ccccc}
  \hline\hline
  \textbf{Function} & \textbf{Average} & \textbf{Std Dev} & \textbf{min} & \textbf{max} \\ 
  \hline
  Scaled Jacobian    & 0.93            & 0.13              & 0.0028       & 1            \\ 
  \hline\hline
  \end{tabular}
  \caption{Scaled Jacobian quality of the current mesh.}
  \label{tab:scaled_jacobian}
\end{table}

This final model is imported into \software{COREFORM} \software{CUBIT} \cite{coreformllc_coreform_2024}, a software for geometry manipulation and hexahedral mesh generation that is typically used in conjunction with \software{SPECFEM}. Within \software{CUBIT}, the \software{SCULPT} tool was used to create a high-quality mesh of the dolphin head, as shown in \cref{fig:mesh}. \software{SCULPT} employs an overlay-grid method, specifically an inside-out approach, for semi-automatic hexahedral meshing. This method begins with a base Cartesian grid that encompasses the geometry and then carves, or sculpts, the geometric features from this grid to produce the final hexahedral mesh \cite{owen_parallel_2014a}: see \cref{fig:mesh_a}. Moreover, the results from the boolean fraction operations may leave some traces, as seen in \cref{fig:mesh_b}. The fragmented geometry tells \software{SCULPT} explicitly where to mesh in such complex shapes. \Cref{fig:mesh_b} shows a small deformation in the outer part of the bone. This deformation, not present in the geometry, can be clearly seen as a line on the surface of the right jaw. This is an effect of \software{SCULPT} trying to create a conformal mesh for the acoustic fat and the jaw at the same time. 

It should be noted that \software{BLENDER}, \software{MESHLAB}, and \software{CUBIT} each incorporate their own tools for cleaning and manipulating geometry through Boolean operations, which are necessary for \software{SCULPT} to create complex meshes with mesh adaptivity. As previously mentioned, \software{FREECAD} was better suited to our needs because it ensures these operations produce a fragmented geometry mesh, i.e., a geometry where the intersections between objects have conformal surface meshes.

\software{CUBIT} allows us to define the material properties of each anatomical feature, assign faces to boundary conditions, and export the resulting mesh directly to \software{SPECFEM3D} through \software{PYTHON} functions included in the package. Although simple to use, \software{SCULPT} can yield different results depending on the parameters used. For example, the present model was successfully meshed with \software{SCULPT} using a maximum cell size of 2.5 mm but it may produce distorted elements if the cell size is decreased. This may seem counter-intuitive and is largely due to the nature of mesh sculpting and mesh adaptivity. At this stage, some trial and error is necessary to determine the appropriate parameters. Most of these mesh-related problems can, in fact, be mitigated by drastically reducing the mesh size, for example, to 1 mm. However, this approach can lead to an exponential increase in the number of elements and, consequently, the computational involved costs. 

According to the documentation provided with \software{CUBIT}, the mesh quality, as measured by the scaled Jacobian, should ideally exceed 0.2, although the method can function with values greater than 0. The scaled Jacobian measures the element quality based on the determinant of the Jacobian matrix \cite{knupp_algebraic_2001}. The largest element size was 2.5 mm, covering the majority of the mesh volume (\cref{tab:scaled_jacobian}). At interfaces and/or particularly complex features, such as jaws and acoustic fats (e.g. \cref{fig:mesh_b,fig:mesh_c}), mesh adaptivity was employed with one layer using the material mesh adaptivity feature in \software{SCULPT} and two Laplacian smoothing iterations \cite{owen_parallel_2012}. Additional parameters are specified in \cref{sec:appendix}. Using 24 processors, the \software{SCULPT} tool took about 24 minutes to generate a mesh consisting of 16 million elements and consuming approximately 100 GB of RAM. The output mesh has a maximum element side size of 2.5 mm, by construction, and minimum element side size of 0.1 mm determined automatically by \software{SCULPT} when performing adaptivity meshes to handle small complex shapes.

Finally, alternative methods and algorithms for automatic hexahedral meshing could be taken into consideration. This is a critical step in using SEM as wave propagation solver. For example, it is possible to generate hexahedral meshes from tetrahedral meshes \cite{gao_improved_2025}. However, the quality of the resulting meshes is generally poor \cite{baker_mesh_2005}. Improving such methods may alleviate the efforts required by hexahedral meshing for complex geometries.

\subsection{\label{subsec:simulation}Simulation setup}

\begin{figure}
    \subfloat[\label{fig:plane_wave_time}]{
        \includegraphics[width=0.25\textwidth, trim= 0 50 50 0, clip]{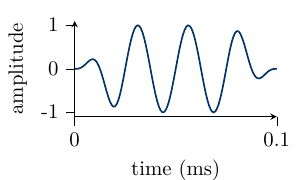}
    }
    \subfloat[\label{fig:plane_wave_spectre}]{
        \includegraphics[width=0.25\textwidth, trim= 0 50 50 0, clip]{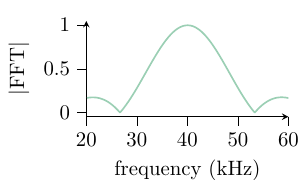}
    }
    \subfloat[\label{fig:plane_wave_tappered}]{
         \includegraphics[width=0.4\textwidth, trim= 0 35 0 30, clip]{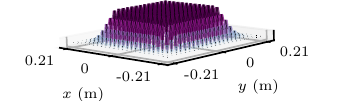}
    }
    \caption{\label{fig:plane_wave}a) Source signal with 4-cycles at 40 kHz. The waveform has been windowed with a Tukey window of length 0.1 ms. b) Spectrum of the signal. The bandwidth of the first zeros (below 30 kHz and above 50 kHz) correspond to length of the window. c) Tapered wave profile. The arrows colors and lengths illustrate the assigned force amplitudes of the signal and point in the -$z$ direction, toward the dolphin's head. }
\end{figure}

Our simulations were performed with \software{SPECFEM3D} version 4.0.0, which supports direct visualization of models and simulations in \software{PARAVIEW} \cite{ahrens_paraview_2005} without requiring additional post-processing. They were carried out on a local cluster with 24 cores, each supporting two threads and a total RAM of 500 GB. They were ran on our mesh of 16 million elements, each element parametrized with 3 Gauss-Lobatto-Legendre (GLL) points ($NGLL=3$). More GLL points could be implemented, but this would increase simulation time significantly. Increasing the $NGLL$ value allows for fewer elements per wavelength, typically 3 to 5 \cite{marty_transcranial_2024}, while maintaining accuracy with larger mesh sizes.
The time step $\Delta t$ is determined by the Courant number $C$, the characteristic element size $h$, and the wave speed $v$. Both $\Delta t$ and $C$ are automatically calculated by \software{SPECFEM3D} and must satisfy $\Delta t<C \min (h/v)$ \cite{komatitsch_spectral_1998}. For this simulation, $\Delta t = 9.5 \times 10^{-9}$ seconds and the total simulation time is $t=0.7$ ms.
Boundary conditions were defined using Clayton-Engquist absorbing boundary conditions \cite{komatitsch_introduction_1999}, which minimize reflections at the mesh boundaries. Also known as ``Stacey'' conditions, they are simple to implement in \software{SPECFEM3D} but may produce strong reflections at grazing angles. A tapered profile was accordingly applied near each border (see \cref{fig:plane_wave_tappered}). This profile gradually reduces the wave amplitude near the edges using a smooth Gaussian function, which helps to absorb outgoing waves and minimize artificial reflections. 

Through the rest of the present study, we shall discuss the results of two simulations: a plane-wave incident on the head and a point source within the head. Each simulation was run in parallel on 24 processors, using about 50 GB of RAM and lasting 45 hours of real time on our local cluster.
\software{SPECFEM3D} supports GPU acceleration \cite{komatitsch_fluid_2011}, which could reduce the computation time by a factor of 5 to 10; its scalability on large clusters is very straightforward, and the simulations presented here could be completed within minutes on a typical ``supercomputer''.
    
The first simulation consists on a plane-wave source placed in front of the dolphin head, emitting four cycles with a fundamental frequency $f_0=$ 40 kHz, and Tukey-windowed as per \cref{fig:plane_wave_time,fig:plane_wave_spectre}. In practice, multiple point sources fired simultaneously along the $z$ direction towards the dolphin head. Approaching the boundaries of the numerical domain, a Gaussian-shaped taper (decaying with decreasing distance from the boundary) was applied to the magnitude of point sources (\cref{fig:plane_wave_tappered}). The resulting wavefield was recorded near the approximate locations of the left and right Tympano-Periotic Complexes (TPC). The TPC is a complex structure that includes the cochlea, vestibule, and semicircular canals, and is responsible for hearing and balance in dolphins \cite{cozzi_chapter_2017}. Our CT scan data and mesh do not honor the extreme complexity of this area, but the simulations should describe fairly accurately vibrations at the end of the acoustic-fat area.

The second simulation involves a single point source inside the dolphin head. The source was placed near the location of the phonic lips, responsible for emitting click in odontocetes \cite{cranford_observation_2011}. The source was defined as a force oriented along the $z$ axis, varying in time as in \cref{fig:plane_wave_time}. 

\subsection{\label{subsec:materials}Material properties}
The density and shear and compressional velocities of the dolphin head were derived from the CT scan described above, via the HU values. On the basis of HU values, we identified four different types of tissue and assigned a single set of average density and shear and compressional velocity to each, as per \cref{tab:material_properties}. Other authors prefer to derive more complex 3D density-velocity maps directly from the HU values \cite{wei_acoustic_2015}; this could make a significant difference as the mechanic properties of some tissues (e.g., the melon) are known to vary, but it is not relevant to the feasibility of our general approach. 

\begin{table}[!ht]
  \begin{tabular}{cccccc}
    \hline\hline
    \textbf{Material} & $\bm{\mathrm{HU_{min}}}$ & $\bm{\mathrm{HU_{max}}}$ & \boldmath$\rho$& \boldmath$v_p$ & \boldmath$v_s$ \\ \hline
    Soft tissue     & -35        &  110   & 1013  & 1536  & 215   \\ 
    Acoustic fat    & -115       &  -35   & 928   & 1390  & 186   \\ 
    Melon           & -115       & -35    & 884   & 1316  & 184   \\ 
    Bone            & 235        &  2030  & 2035  & 3400  & 1817  \\ 
    Water           & $<$-2000   &        & 1028  & 1480  & 0   \\ 
    \hline\hline
  \end{tabular}
  \caption{Material properties of the dolphin head and surrounding medium, including mean Hounsfield units (HU), volumetric density $\rho$ (kg$\cdot$m$^{-3}$), compressional wave velocity $v_p$ (m$\cdot$s$^{-1}$) and shear wave velocity $v_s$ (m$\cdot$s$^{-1}$).}
  \label{tab:material_properties}
\end{table}

\subsection{\label{subsec:validation}Model Validation}
We approximately quantify the numerical error associated with our model via a simple yet mathematically precise test based on the so-called reciprocity theorem \cite{aki_quantitative_1980}. Consider the Green's function $G=G(\r, t|\r_1, \tau)$ solution of $\mathcal{L}G=\delta(\r-\r_1)\delta(t-\tau)$, where the differential operator $\mathcal{L}$ describes as always the propagation of acoustic or elastic waves. For any pair of locations $\r_1, \r_2$, the reciprocity theorem states that

\begin{equation}
    G(\r_1, t|\r_2,\tau) = G(\r_2, t|\r_1,\tau),
\end{equation}
i.e., the physical quantity measured at point $\r_2$ at time $\tau$ due to an impulse source at $\r_1$  at time $t$ is equal to that measured at $\r_1$ at time $t$ due to an impulse source at $\r_2$ at $\tau$. This reflects spatial symmetry: i.e., swapping the locations of source and receiver results in the same recorded signal.

\begin{figure}
    \centering
    \subfloat[\label{fig:reciprocity_sources}]{
        \includegraphics[width=0.3\textwidth]{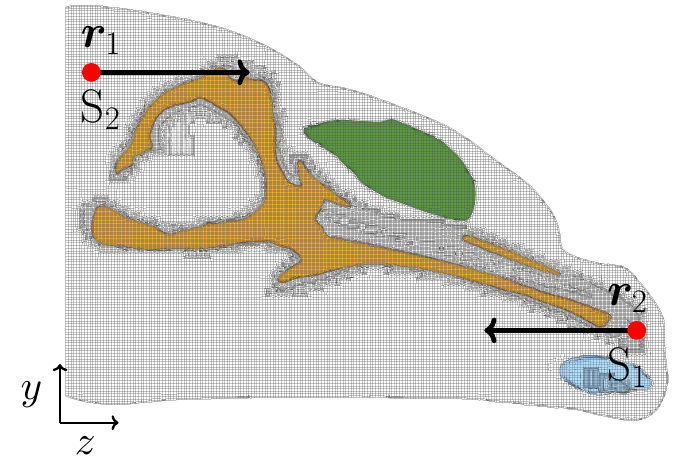}
    }
    \subfloat[\label{fig:reciprocity_velocity}]{
        \includegraphics[width=0.3\textwidth]{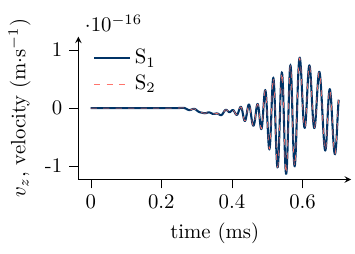}
    }
    \subfloat[\label{fig:reciprocity_difference}]{
        \includegraphics[width=0.3\textwidth]{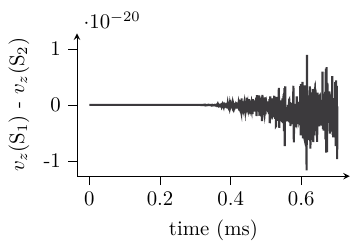}
    }
    \caption{\label{fig:reciprocity} Reciprocity Theorem.  Location of the sources for the test, shown on a slice of the head mesh along the $yz$-plane at $x=0$. b) Comparison of recorded signals. The recorded signal $\mathrm{S_1}$ fired from $\r_1$ is ``equal'' to the recorded signal $\mathrm{S_2}$ fired form $\r_2$. c) Absolute difference between the two signals. The difference is four orders of magnitude smaller than the signal themselves.}
\end{figure}

We place two sources on the (approximately) median plane of the dolphin head, parallel to one another and pointing in opposite directions as depicted in \cref{fig:reciprocity_sources}. \Cref{fig:reciprocity_velocity} shows the $z$-components of the velocity field generated by a point source at $\r_1$ and recorded at $\mathrm{S}_1$, compared with the velocity at $\mathrm{S}_2$ from a source at $\r_2$. Their difference, which should be exactly zero according to theory, is four orders of magnitude (i.e. 40 dB) smaller than the computed signals themselves. We consider this amount of numerical noise to be very small, and certainly small enough for our current goals. 

\section{\label{sec:results}Results}

\subsection{\label{subsec:plane_wave}Incoming plane wave}

\begin{figure}
    \centering
    \subfloat[\label{fig:sim_t1}t=0.17 ms]{
       \includegraphics[width=\textwidth/2]{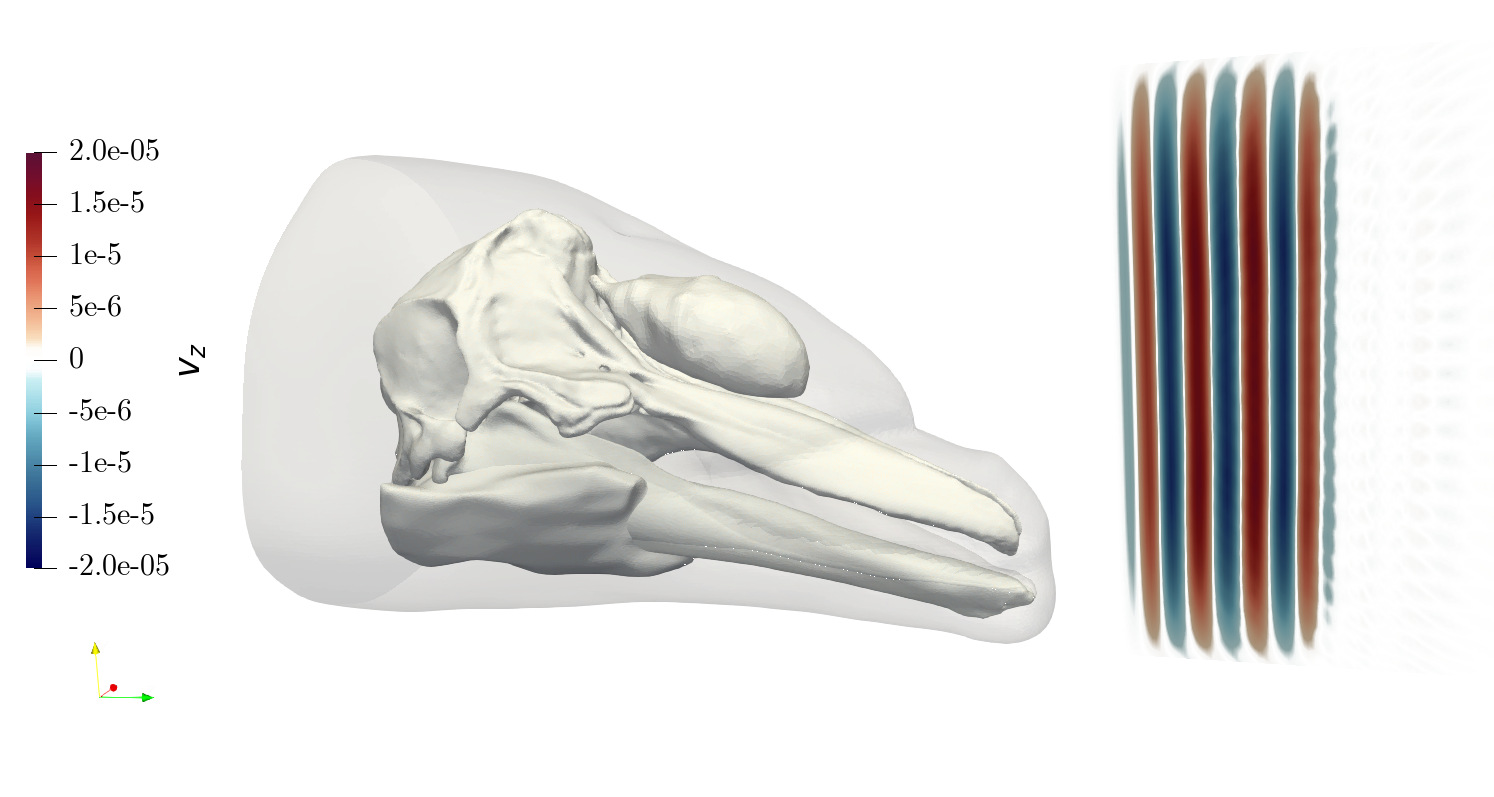}}
    \subfloat[\label{fig:sim_t2}t=0.32 ms]{
        \includegraphics[width=\textwidth/2]{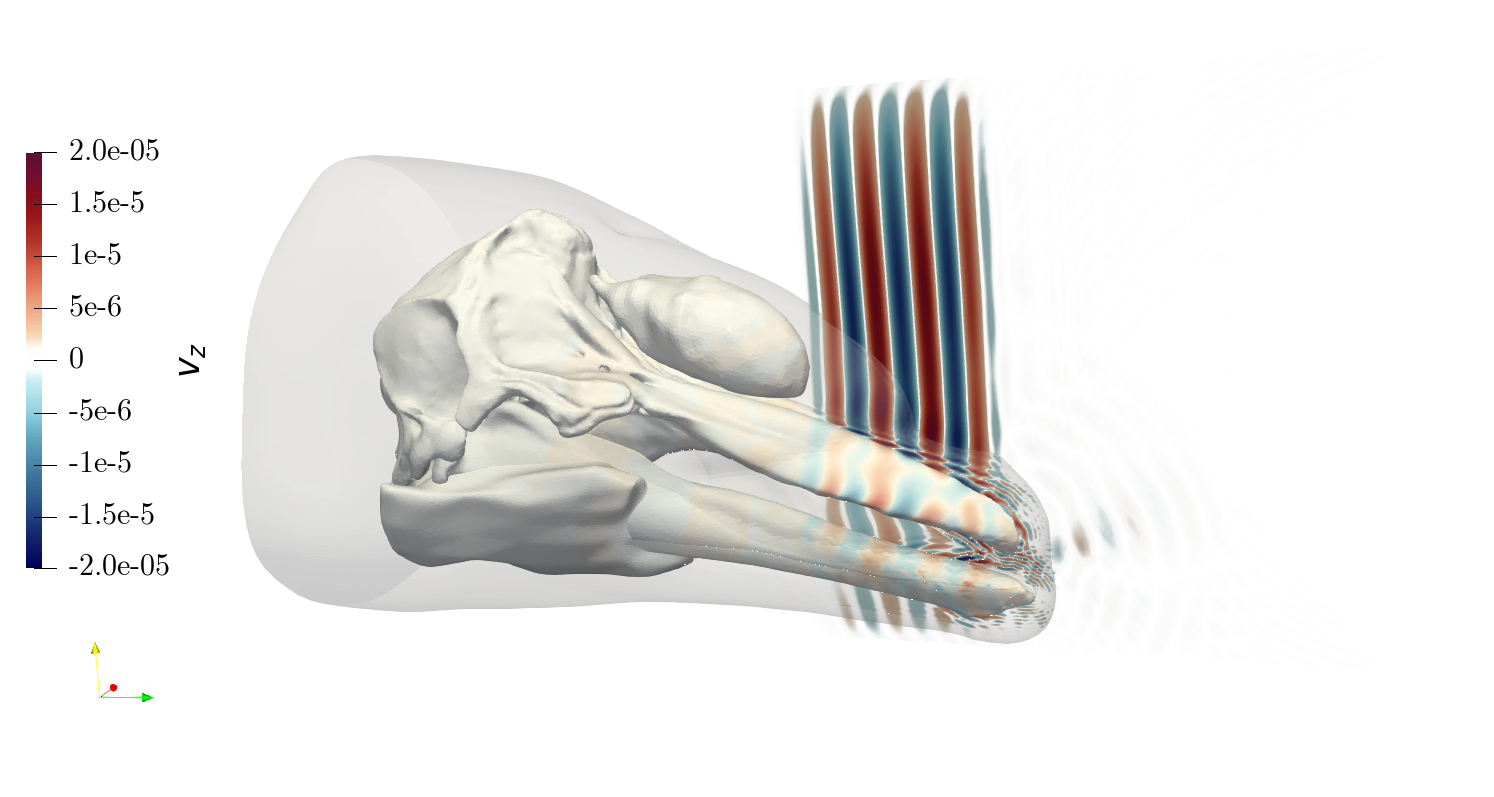}}\\
    \subfloat[\label{fig:sim_t3}t=0.46 ms]{
        \includegraphics[width=\textwidth/2]{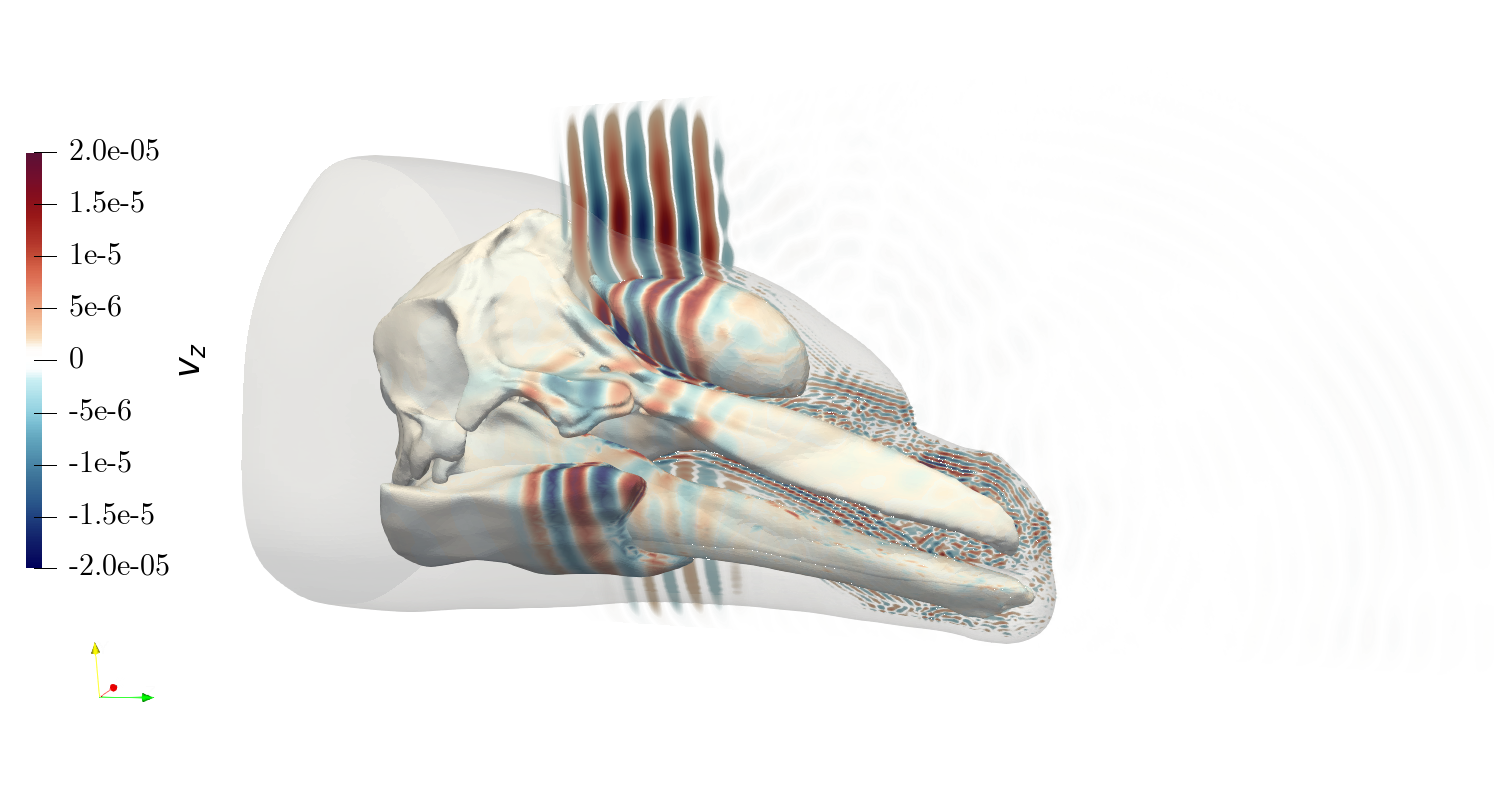}}
    \subfloat[\label{fig:sim_t4}t=0.59 ms]{
        \includegraphics[width=\textwidth/2]{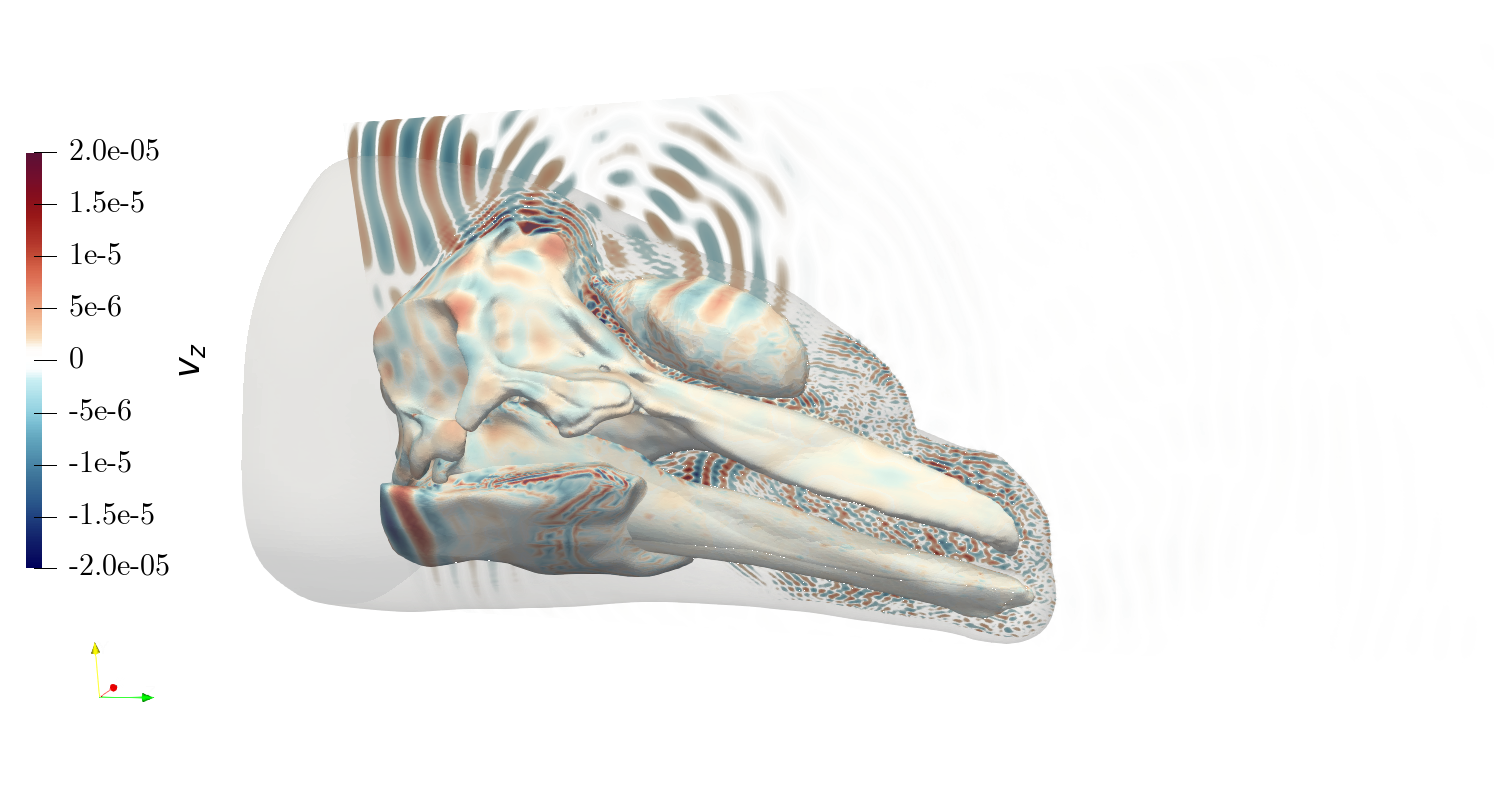}}
    \caption{\label{fig:sim}A slice along the $yz$ plane at $x=0$ is shown depicting the wavefront propagating in the fluid domain towards the dolphin head and through its different tissues at different times $t$. The colorbar scale show the amplitude of the $z$ component of the particle velocity $v_z$. This value was projected onto the surface of each tissue to highlight the wave-propagation phenomenon. The scale across all 3D tissues was adjusted uniformly to enhance visualization.}  
\end{figure}

\begin{figure}
    \subfloat[\label{fig:at_ears_pressure}]{
        \includegraphics[width=0.3\textwidth]{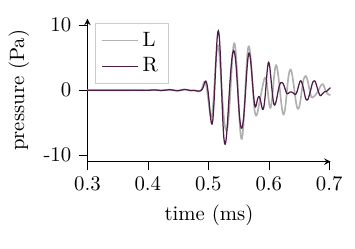}
    }
    \subfloat[\label{fig:at_ears_spectrum}]{
        \includegraphics[width=0.3\textwidth]{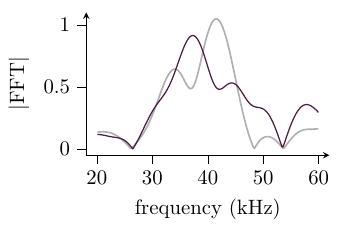}
    }
    \subfloat[\label{fig:at_ears_velocity}]{
        \includegraphics[width=0.3\textwidth]{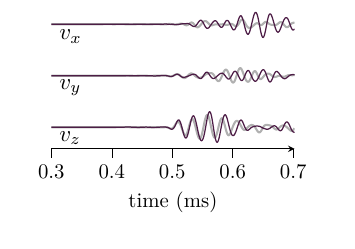}
    }
    \caption{\label{fig:at_ears}Recorded signals near the Tympano-Periotic Complex (TPC), at both left (L, depicted in gray thick lines) and right (R, in violet thin lines) ears. a) Pressure, b) Pressure spectrum. c) Each component of the particle velocity. The $v_z$ component consists mainly of the direct signal with some reverberation. However, in the other components, the reverberations are more prominent. }
\end{figure}

We made wide use of the \software{PARAVIEW} software to visualize all our simulation results. Data from \software{SPECFEM} were imported directly into \software{PARAVIEW} as XDMF (eXtensible Data Model and Format) and associated  HDF5 (Hierarchical Data Format) files. Then, each STL file associated to a different anatomical feature (tissue type) was imported and the \software{Point} \software{Dataset} \software{Interpolator} module was used to project the volume solution onto the outer surface of each feature. A slice along the median $yz$-plane at $x=0$ shows the wave propagation towards and into the dolphin head in \crefrange{fig:sim_t1}{fig:sim_t4}. The colorbar scale corresponds to the particle velocity $v_z$ on the same plane. For illustration purposes, the amplitude in all 3D-displayed materials was increased uniformly to match that of the surrounding water.

The impedance mismatch between different tissues and water is clearly visible in \cref{fig:sim_t3}, where the amplitude of the particle velocity is greater in the acoustic fats than in the jaw. Compared to the simple source signal that was injected into the model, the plane-wave simulation in \crefrange{fig:sim_t2}{fig:sim_t4} shows a very complex wavefield, dominated by the effects of the many sharp discontinuities characterising the dolphin anatomy, including guided waves and reverberations. 
In \cref{fig:sim_t3,fig:sim_t4}, the high-frequency reflections observed near the skull are likely visual artifacts, as the cross-sectional view (2D slice) captures waves that have presumably been reflected in various directions near the dolphin’s beak, giving the impression that the reflections are occurring at the skull. 
The frequency spectrum (\cref{fig:at_ears_spectrum}) of the recorded signal is largest around 40 kHz, and has its absolute maximum slightly below 40 kHz at the left ear, and slightly above at the right ear. Spectral amplitudes decay to zero approximately in the same way as those of the source signal (\cref{fig:plane_wave_spectre}),  with the bandwidth between zeros corresponding to the length of the Tukey window.
In \cref{fig:at_ears_pressure,fig:at_ears_velocity}, both pressure and particle velocity exhibit a slightly modulated source signal followed by a complex, high-amplitude reverberation pattern, consistent with the qualitative findings of \citet{hejazinooghabi_contribution_2021}.

The results of our relatively simple simulation show clearly that SEM is a powerful tool to capture the many complexities of wave propagation through the anatomy of marine mammals. In future work, and via a much more detailed analysis of numerical data, this can contribute to understanding the mechanisms underlying their auditory and biosonar systems.

\subsection{\label{subsec:point_source}Outgoing point source}

\begin{figure}
    \centering
    \subfloat[\label{fig:sim_out_t1}t=0.02 ms]{
       \includegraphics[width=\textwidth/2]{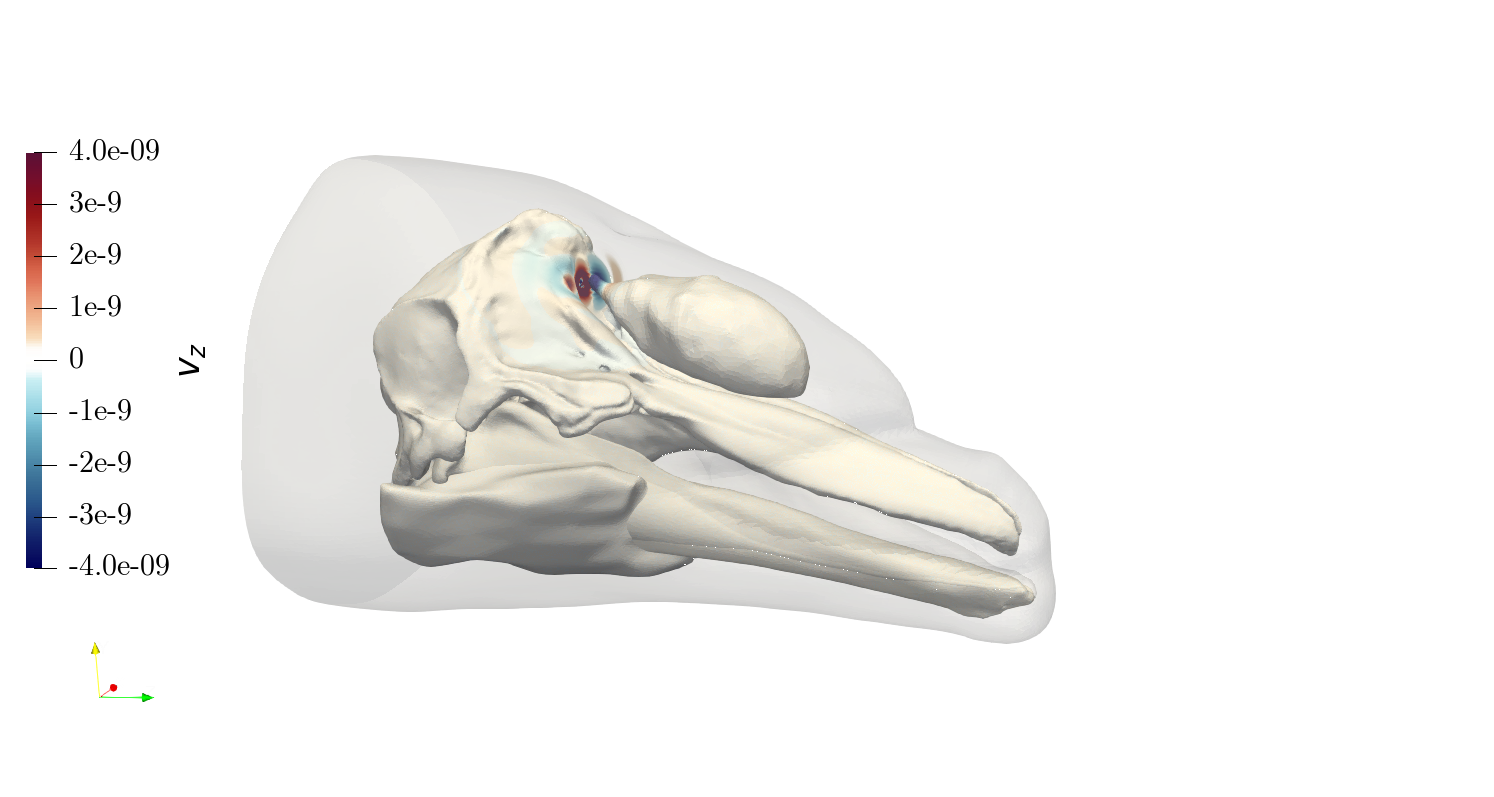}}
    \subfloat[\label{fig:sim_out_t2}t=0.15 ms]{
        \includegraphics[width=\textwidth/2]{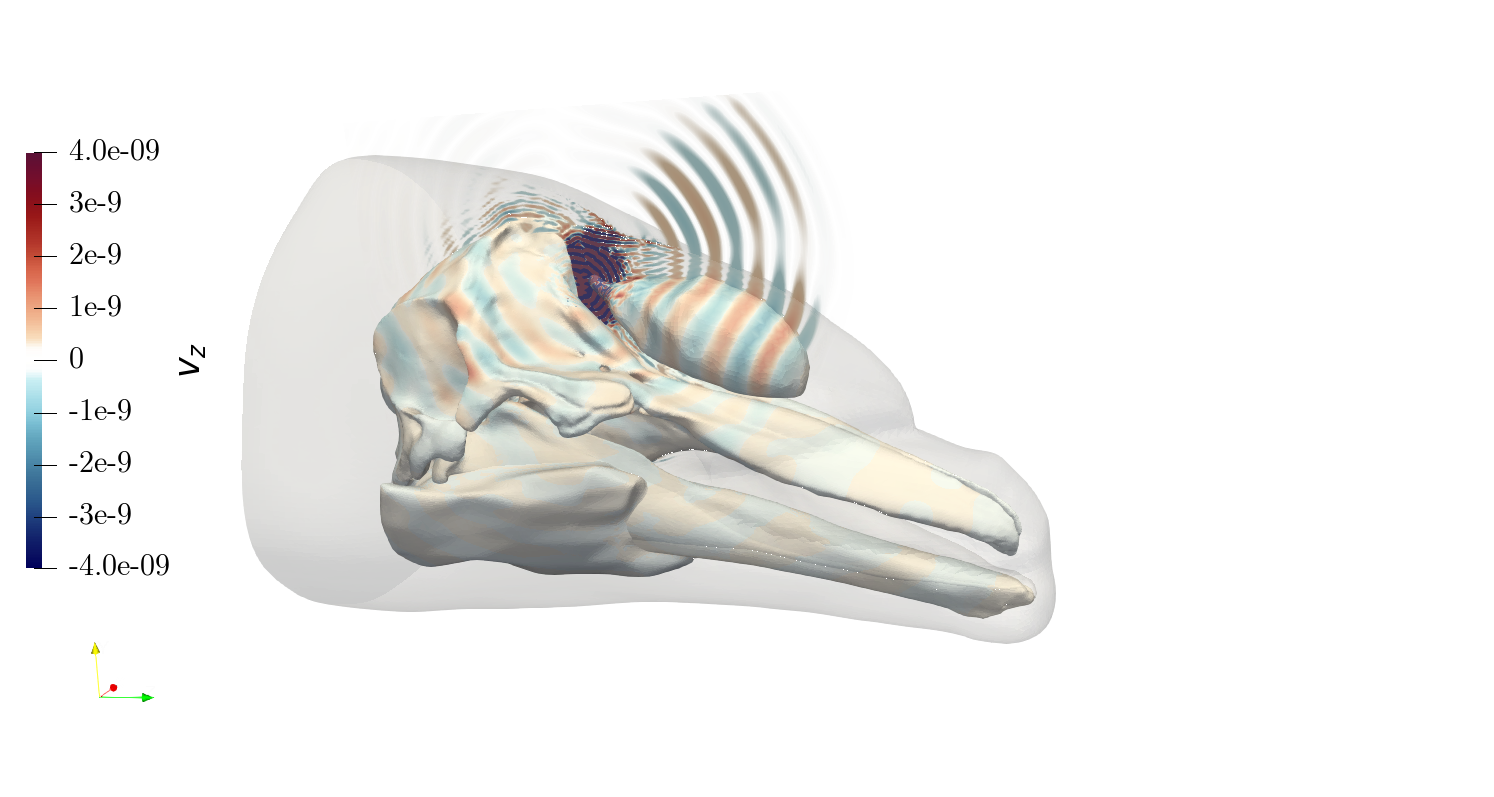}}\\
    \subfloat[\label{fig:sim_out_t3}t=0.29 ms]{
        \includegraphics[width=\textwidth/2]{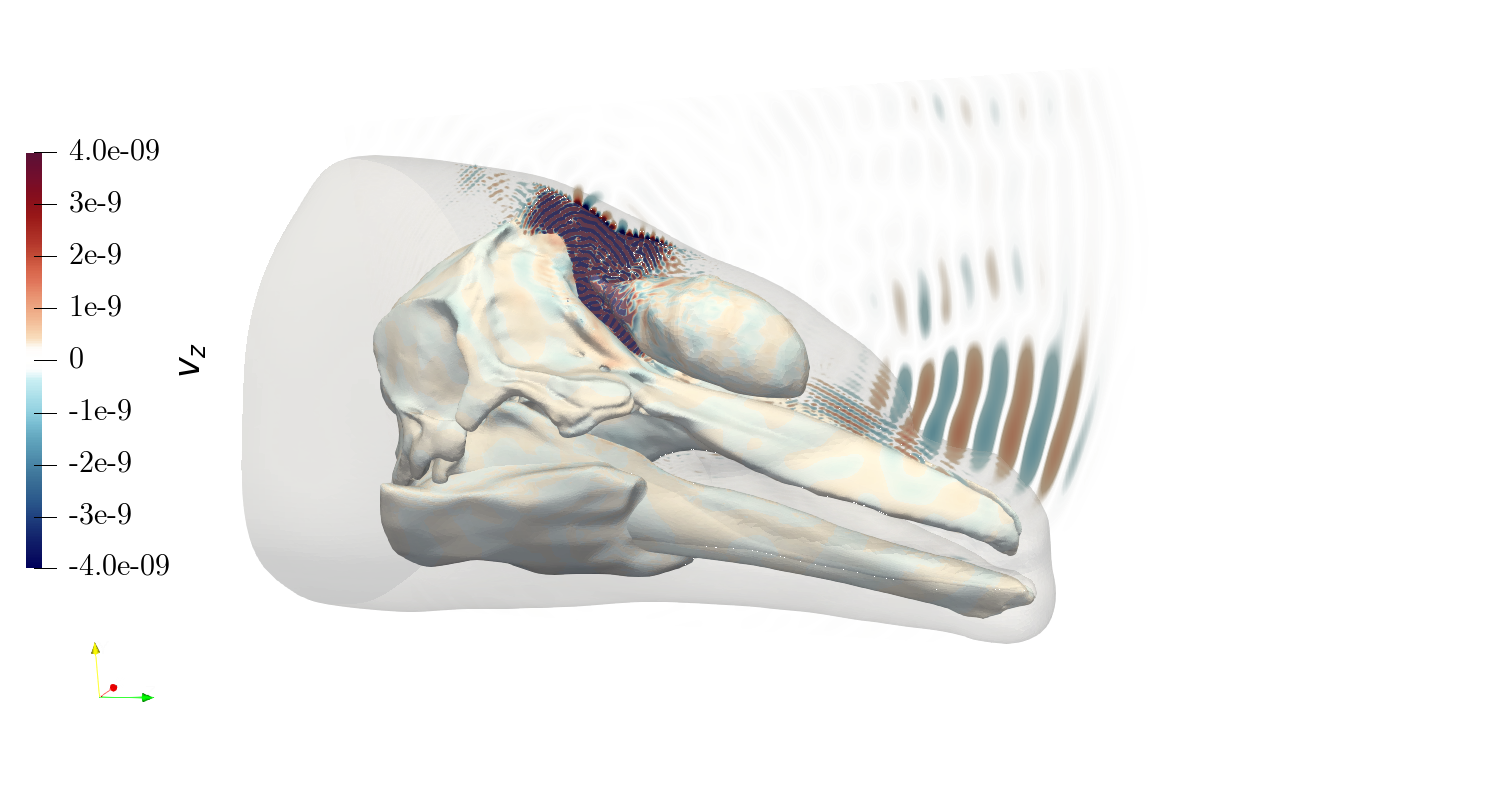}}
    \subfloat[\label{fig:sim_out_t4}t=0.41 ms]{
        \includegraphics[width=\textwidth/2]{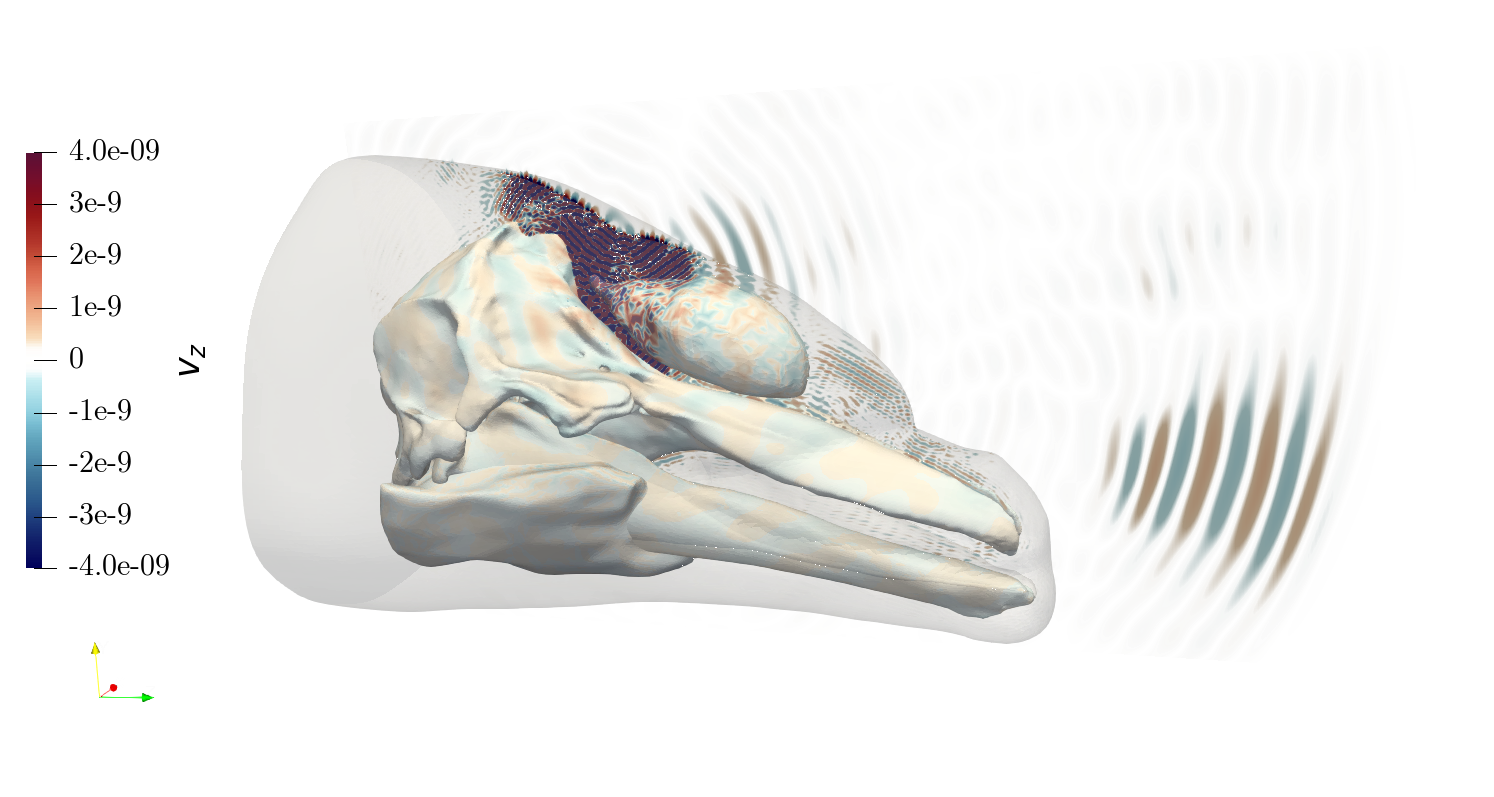}}
    \caption{\label{fig:sim_out}Outgoing click propagation. The color scale is defined so as to be able to visualize the relative-low amplitude compressional wave propagating through water. This results in the saturation of vibration amplitude, within the head, in the immediate vicinity of the source. The scale show the amplitude of the $z$ component of the particle velocity $v_z$. This value was projected onto the surface of each tissue to highlight the wave-propagation phenomenon. The scale across all 3D tissues was adjusted uniformly to enhance visualization.}  
\end{figure}

\Cref{fig:sim_out} shows the wavefield originated by a point source at the approximate location of the phonic lips, responsible for generating dolphin clicks. We used the same synthetic signal shown in \cref{fig:plane_wave_time} for illustration purposes, although the frequency range of dolphin clicks is typically higher (around 120 kHz for \textit{Tursiops Truncatus}). Note that the pathway followed by the outgoing signal is aligned with the melon, as seen in \cref{fig:sim_out_t3,fig:sim_out_t4}. \Cref{fig:sim_out_t2} also shows the primary wave propagating at an angle, above the head and slightly upwards; the late wavelet propagating above the head in \cref{fig:sim_out_t4} appears to be a reflection of that, from the skull itself. This reflection may disappear or become less prominent if air is included in the model, which we have so far omitted. This is a non-trivial issue, as air was largely absent from the CT scan.

\section{\label{sec:discussion}Discussion and outlook}

We demonstrated that the Spectral-Element Method can be used to successfully simulate 3D wave propagation through a bottlenose dolphin head. Ours was essentially a feasibility test, conducted on relatively limited computational infrastructure, but enough to infer that, after migrating our code to a large high-performance cluster, we shall be able to model most of the complexities in the anatomy of real dolphins. Importantly, this study is the first instance of using SEM in this context and one of only relatively few attempts at modeling 3D time-domain wave propagation through anatomical structures \cite{hejazinooghabi_contribution_2021,marty_transcranial_2024,wei_validated_2024,bachmann_source_2020}.

Features that were omitted in this study, but might turn out to play a significant role, include the dolphin's teeth and air sacs and ducts \cite{au_sonar_1993,dobbins_dolphin_2007}. We also assumed homogeneous tissue properties within individual anatomical features, which may have a significant effect e.g. on wave focusing/defocusing. The rationale for sticking to a simplified model is twofold: on the one hand, the specimen available to us was an old dolphin in relatively bad shape; some teeth were missing, soft tissues around the beak were deformed, the majority of the air spaces were missing in the CT scan. A lot of reshaping and smoothing was, accordingly, needed. On the other hand, the difficulty of hexahedral meshing is well known, and grows with the complexity of the structures that are to be meshed. Semi-automatic meshers like \software{SCULPT} alleviate this process, but still, before we came up with a good mesh, and with a reliable protocol (described in much detail in \crefrange{subsec:ct_scans}{subsec:meshing}) to derive it, a lot of trial and error  was necessary. The jaws and the acoustic fats, for instance, were particularly problematic. 
Since \software{SCULPT} can only deal with geometries that are conformal and fragmented at interfaces for external STL files and adaptative meshes, \software{CUBIT} often failed and crashed, as we fed it multiple, trial versions of those features, without providing information. We were surprised to find that \software{FREECAD} could succeed where all other ``geometry tools'' that we tested, such as \software{BLENDER} or \software{MESHLAB}, failed. Since these are all open-source packages, it is possible that some plug-ins, of which we are not aware, exist to deal with such task. If that were the case, we would recommend to try and perform all necessary boolean operations directly in \software{BLENDER}, without recurring to \software{MESHLAB} and/or \software{FREECAD}.

One relatively important issue with \software{SCULPT} is that the whole model needs to be meshed at once. Currently,  \software{SCULPT} does not support meshing individual features separately: in our case, for instance, it would have been preferable to mesh the acoustic fats first, then the jaws, skull, soft tissues, and finally the water. In general, it would be easier to refine the mesh locally, only where it is needed (e.g., the Tympano-Periotic Complex), before merging individual features in a single, large model.  

\software{SCULPT} will always generate a mesh, but, depending on the input data, the mesh quality measured by its scaled Jacobian can be below zero, thus unusable \cite{igel_computational_2016}. Mesh quality can be significantly affected by very small adjustments in the input data. Even very minor smoothing of anatomical features can compromise it, owing to the iterative process of \software{SCULPT}; in particular, when trying to create the coarsest possible mesh according to the source-signal wavelength, in a medium that includes, like ours, small ducts and sections. All this means that the user has to systematically and carefully check the output of \software{SCULPT}, which slows down the process.

Although a maximum mesh-size of 2.5 mm allows for signals up to $f_0=60$ kHz, assuming 10 points per wavelength, and up to $f_0=200$ kHz using 3 points per wavelength, the simulations in this study were limited to relatively low frequencies, with $f_0=40$ kHz. Higher frequencies would imply longer simulation runs, not needed for the purpose of this study. In future work, when extending \software{SPECFEM3D} to a broader frequency range (and more powerful, high-performance computer infrastructure), we also expect to conduct additional validation experiments such as that in \cref{subsec:validation}, and/or direct comparison against other methods such as \software{SALVUS} \cite{marty_transcranial_2024}, \software{COMSOL} \cite{wei_validated_2024} or \software{k-WAVE} \cite{hejazinooghabi_contribution_2021}. 

\software{SPECFEM3D} assumes a Cartesian reference frame, and the boundaries of the simulation domain are aligned with its axes. This constraint arises from the nature of seismology simulations, which typically model only a portion of the domain (the earth), enclosed in a rectangular prism, with zero shear stress prescribed on (only) the upper surface (i.e., the earth's outer surface). While this setup is well-suited for many geophysical applications, it imposes a fixed geometry on the domain boundaries. In contrast, other high-order numerical models mesh the outer surface as tightly as possible, sometimes following a smooth version of the the external surface, namely a ``convex hull'' \cite[Fig. 9]{beriot_automatic_2021}. Adopting a similar approach could help reduce computational costs, but implementing it in \software{SPECFEM3D} may be challenging.

By assuming homogeneous, isotropic acoustic properties for all tissues independently, we also overlook known gradients in stiffness and lipid composition, for example within the melon \cite{norris_sound_1974}. Future work may integrate 3D variations in material parameters as well as anisotropy, ideally based on empirical measurements. Likewise, attenuation (which \software{SPECFEM3D} can honor) will be incorporated in the model by taking into account viscoacoustic-viscoelastic coupling.

Finally, \cref{fig:sim_out} shows high-amplitude vibrations and multiple reverberations near the source. Both effects might be strongly reduced once our model will account for air spaces with the dolphin head, and for attenuation. Air spaces, particularly those located near the phonic lips, e.g. the nasal passages or \textit{choanae} \cite{cozzi_chapter_2017}, might help to absorb these vibrations, preventing them to reach the skull. They may also enhance the amplitude of the outgoing signal, acting as a baffle \cite{wei_finite_2018}. By incorporating attenuation, high-frequency reverberations will decrease faster in time than the fundamental frequency of the original signal. 

We submit that the Spectral-Element Method may be the most powerful tool available to account for and analyze the effects of these and other complex features of marine mammal anatomy.

\section{\label{sec:conclusion}Conclusions}

We demonstrated that the Spectral-Element Method (SEM) effectively simulates 3D wave propagation in marine mammals, using a detailed dolphin model built from CT scans. By leveraging STL files we simplified the meshing process, turning it into a practical 3D modeling task. We prepared the model carefully and used  \software{SCULPT} for adaptive meshing to capture complex structures accurately.
With \software{SPECFEM3D}, we simulated both plane waves and point sources, unlocking the ability to further explore key issues like dolphin biosonar. 
This approach has wide-ranging applications: it could be used, e.g., to assess how anthropogenic noise affects marine life, recreate behavioural experiments for understanding image formation from sound in dolphins, and shed light on the biophysics of hearing and click generation in marine mammals.
Our work shows the power of computational tools in marine biology, offering a clear path to study and protect marine mammals amid growing environmental pressures. Several hypotheses and well established theories have been proposed to explain, for instance, the sound transmission through the head \cite{norris_sound_1974}, the role of left-right asymmetry \cite{ryabov_role_2023}, the role of mental foramens \cite{ryabov_role_2010}, the two-sonar source theory \cite{cranford_observation_2011} and, in general, the mechanisms of sound reception and conduction \cite{ryabov_mechanisms_2014} in different dolphins. We might be at the point where we can start testing all these theories efficiently.

\section*{Acknowledgments}\label{sec:acknowledgments}
We thank Sandro Mazzariol, Steffen De Vreese, Jean-Marie Graïc, Ksenia Orekhova, and the DIAPHONIA network for their constant feedback on the anatomy and hearing of dolphins and preparing the STL files. We are also grateful to Quentin Grimal for his ongoing involvement and valuable comments, Vadim Monteiller for assisting with the installation of the appropriate version of \software{SPECFEM}, Daniel Peter for his earlier guidance, and Norbert Hofbauer for his helpful tips on using \software{CUBIT}.

We acknowledge financial support under the National Recovery and Resilience Plan (NRRP), Mission 4, Component 2, Investment 1.1, Call for tender No. 1409 published on 14.9.2022 by the Italian Ministry of University and Research (MUR), funded by the European Union -  NextGenerationEU - project title: SWIM: aSsessing the Impact of offshore Wind turbines on Marine mammals in the Adriatic sea - CUP C53D23010230001- Grant Assignment Decree No. 1388 adopted on 01.09.2023 by the Italian Ministry of University and Research (MUR).

\section{Meshing Commands}
\label{sec:appendix}

The detailed commands used for meshing with \software{SCULPT} shown in \cref{tab:sculpt}.
\begin{table*}[bpt!]
\resizebox{.9\textwidth}{!}{%
  \begin{tabular}{ccl}
        \hline\hline
        Command & Value & Detail\\
        \hline 
        \sculpt{processors} & 24 & Sets the number of processors for parallel computations to 24. \\
        \sculpt{size}       & 0.0025 & Specifies the absolute cell size for the Cartesian grid as 0.0025. \\
        \sculpt{box expand} & 0 & Controls the bounding box expansion; 0 means no expansion. \\
        \sculpt{csmooth}    & 6 & Sets the curve smoothing method to linear. \\
        \sculpt{curve\_opt\_thresh} & 0.61 & Sets the minimum metric threshold for curve optimization to 0.61. \\
        \sculpt{smooth}     & 3 & Applies smoothing without surface projections. \\
        \sculpt{max\_opt\_iters}  & 80 & Sets the maximum number of optimization iterations to 80. \\
        \sculpt{max\_pcol\_iters} & 101 & Sets the maximum number of parallel coloring iterations to 101. \\
        \sculpt{pcol\_threshold}  & 0.3 & Sets the parallel coloring threshold to 0.3. \\
        \sculpt{defeature}  & 1 & Enables automatic defeaturing by filtering small features. \\
        \sculpt{thicken\_material } &3 2 0.1 4 0.05 6 0.05 & Expands materials 3, 4 and 6 (skull and acoustic fats) into \\[-.2em]
                                & & surrounding cells with given parameters.\\
        \sculpt{adapt\_type }&8 & Sets the adaptive meshing type to material-based adaptation. \\
        \sculpt{adapt\_material} &3 2 1 0.5 0 0 4 1 0.5 0 0 6 1 0.5 0 0 & Specifies adaptive meshing parameters for materials 3, 4, and 6. \\
    \hline\hline
  \end{tabular}}
  \caption{\label{tab:sculpt}Commands used in \protect\software{SCULPT} for parallel meshing.}
\end{table*}

\bibliographystyle{elsarticle-num-names}        
\bibliography{references}
\end{document}